\documentclass{aastex61}

\received{November 15, 2017}
\revised{December 16, 2017}
\accepted{January 17, 2018}
\submitjournal{ApJ}

\shorttitle{Lupus transition disks}
\shortauthors{van der Marel et al.}

\begin{document}

\title{New insights into the nature of transition disks from a complete disk survey of the Lupus star forming region}

\correspondingauthor{Nienke van der Marel}
\email{astro@nienkevandermarel.com}

\author{Nienke van der Marel}
\affil{Institute for Astronomy, 
University of Hawaii, 
2680 Woodlawn dr.,
96822 Honolulu HI,
USA}
\nocollaboration

\author{Jonathan P. Williams}
\affil{Institute for Astronomy, 
University of Hawaii, 
2680 Woodlawn dr.,
96822 Honolulu HI,
USA}
\nocollaboration

\author{Megan Ansdell}
\affiliation{Department of Astronomy, University of California Berkeley, 501 Campbell Hall \#3411, Berkeley, CA, USA}
\nocollaboration

\author{Carlo F. Manara}
\affiliation{European Southern Observatory,Karl-Schwarzschild-Strasse 2, 85748, Garching bei Munchen, Germany}
\nocollaboration

\author{Anna Miotello}
\affiliation{European Southern Observatory,Karl-Schwarzschild-Strasse 2, 85748, Garching bei Munchen, Germany}
\nocollaboration

\author{Marco Tazzari}
\affiliation{Institute of Astronomy, University of Cambridge, Madingley Road, CB3 0HA,  Cambridge, UK}
\nocollaboration

\author{Leonardo Testi}
\affiliation{European Southern Observatory,Karl-Schwarzschild-Strasse 2, 85748, Garching bei Munchen, Germany}
\nocollaboration

\author{Michiel Hogerheijde}
\affiliation{Leiden Observatory, Niels Bohrweg 2, 2333 CA Leiden, the Netherlands}
\affiliation{Anton Pannekoek Institute for Astronomy, University of Amsterdam, Science Park 904, 1098 XH, Amsterdam, The Netherlands}
\nocollaboration

\author{Simon Bruderer}
\affiliation{Max Planck Institute fur Extraterrestrische Physik, GiessenbachstrassŸe, 85741, Garching bei Munchen, Germany}
\nocollaboration

\author{Sierk van Terwisga}
\affiliation{Leiden Observatory, Niels Bohrweg 2, 2333 CA Leiden, the Netherlands}
\nocollaboration

\author{Ewine F. van Dishoeck}
\affiliation{Leiden Observatory, Niels Bohrweg 2, 2333 CA Leiden, the Netherlands}
\affiliation{Max Planck Institute fur Extraterrestrische Physik, GiessenbachstrassŸe, 85741, Garching bei Munchen, Germany}
\nocollaboration
       
\begin{abstract}
Transition disks with large dust cavities around young stars are promising targets for studying planet formation. Previous studies have revealed the presence of gas cavities inside the dust cavities hinting at recently formed, giant planets. However, many of these studies are biased towards the brightest disks in the nearby star forming regions, and it is not possible to derive reliable statistics that can be compared with exoplanet populations. We present the analysis of 11 transition disks with large cavities ($\geq$20 AU radius) from a complete disk survey of the Lupus star forming region, using ALMA Band 7 observations at 0.3" (22-30 AU radius) resolution of the 345 GHz continuum,  $^{13}$CO and C$^{18}$O 3--2 observations and the Spectral Energy Distribution of each source. Gas and dust surface density profiles are derived using the physical-chemical modeling code DALI. This is the first study of transition disks of large cavities within a complete disk survey within a star forming region. The dust cavity sizes range from 20-90 AU radius and in three cases, a gas cavity is resolved as well. The deep drops in gas density and large dust cavity sizes are consistent with clearing by giant planets. The fraction of transition disks with large cavities in Lupus is $\gtrsim$11\%, which is inconsistent with exoplanet population studies of giant planets at wide orbits. Furthermore, we present a hypothesis of an evolutionary path for large massive disks evolving into transition disks with large cavities.
\end{abstract}


\section{Introduction}
Planets form in protoplanetary disks of gas and dust around young stars \citep{Armitage2011,WilliamsCieza2011}.
Much observational effort has been devoted to studies of these disks around classical T Tauri stars and to the later debris disk stage when the gas has dissipated \citep{Wyatt2008}. It remains unclear how and when disks evolve in between these two stages. 
Transition disks with cleared out cavities in the inner part of the dust disk are particularly interesting as these disks are likely in the middle of active evolution and possibly planet formation \citep[e.g.][]{Strom1989,Calvet2002,Espaillat2014}.

These disks were originally identified through a dip in the mid-infrared part of the Spectral Energy Distribution (SED) and modeled as axisymmetric dust disks with inner cavities. 
Resolved millimeter interferometry pre-ALMA images \citep[e.g][]{Dutrey2008,Brown2009,Andrews2011,Isella2012} and subsequently at much higher image quality with ALMA \citep[e.g.][]{vanderMarel2013,Perez2014} have revealed the dust rings and cavities. The presence of a cavity suggests that a
companion (either substellar or planetary) may have cleared out its orbit in the disk \citep{LinPapaloizou1979}. However, other mechanisms, such as photoevaporation \citep[e.g][]{Alexander2014} or dead zones \citep[e.g.][]{Turner2014} or a combination of processes \citep{Rosotti2013,Pinilla2016dz} may also be responsible for the presence of gaps. Photoevaporation is often ruled out due to the high accretion rates, which have been detected in a large fraction of transition disks \citep{Najita2007,Najita2015, OwenClarke2012, Manara2014, Espaillat2014, Ercolano2017}. 
For several transition disks, the gas structure inside the dust cavities has been resolved through ALMA CO observations  \citep{Bruderer2014,vanderMarel2015-12co,SPerez2015,Zhang2014,vanderMarel2016-isot,Canovas2016,Dong2017,Boehler2017,Fedele2017}. All disks to date that have been sufficiently resolved show gas cavities that are smaller than the dust cavities, an indication of one or more undetected companions \citep{Zhu2011,Pinilla2012b,Fung2014}. Small gas rings were previously suggested by near infrared observations of the rovibrational CO lines in some transition disks \citep{Pontoppidan2008,Brown2012a,Carmona2017}, but ALMA observations allowed the determination of the radial structure of the gas surface density profile through imaging of the rotational CO line, tracing the bulk of the gas, suggesting the clearing by a companion. A consequence of this clearing process is the existence of pressure bumps at the outer gap gas edge, where millimeter dust will be trapped due to gas-dust drag \citep[e.g.][]{Weidenschilling1977,Zhu2011,Pinilla2012b} and show a narrow dust ring or in certain cases asymmetric dust rings due to Rossby-wave instability \citep{BargeSommeria1995,KlahrHenning1997} or eccentric planet orbits \citep{Ataiee2013,Ragusa2017}. The observed structures in transition disks through multi-wavelength continuum observations are consistent with the dust trapping mechanism \citep{vanderMarel2013,vanderMarel2015vla,Casassus2015,Pinilla2015beta,
Pinilla2017sr24s} in combination with planet clearing. 

Planet candidates have been found in a handful of transition disks \citep[e.g.][]{KrausIreland2012,Quanz2013,Currie2015,Sallum2015}, but for many other disks, only stringent upper limits of a few Jupiter masses can be set \citep[e.g.][]{Maire2017,Pohl2017}, casting doubt on the presence of giant planets inside dust cavities. Also, giant planets at wide orbits are uncommon around main sequence stars ($\sim$0.8\% for all spectral types) \citep{Bowler2016}. On the other hand, most transition disks studies have focused on bright, well-known transition disks around early-type (G and earlier) stars, which may not be representative of the entire transition disk population and cannot be compared statistically with planet formation theory. 

Our Band 7 Lupus ALMA disk survey \citep{Ansdell2016}, a near-complete (96\%) submillimeter study of all protoplanetary disks in the Lupus star forming region, has led to the discovery of several new transition disks, and confirming the presence of a dust cavity for several transition disk candidates based on the modeling of the Spectral Energy Distribution \citep[SED][]{Merin2008,Merin2010,Bustamante2015,vanderMarel2016-spitzer}. In this paper we present the analysis of the radial surface density profile of the dust and gas in 11 transition disks in Lupus, using these ALMA data of the 335 GHz (890 $\mu$m) continuum and the $^{13}$CO and C$^{18}$O 3--2 line observations. We make use of the physical-chemical code DALI \citep{Bruderer2012,Bruderer2013} to extract gas and dust surface density profiles from the line and continuum data. Such a code is necessary to interpret the CO emission, as CO abundances and gas temperatures highly vary throughout the disk due to physical effects (e.g. freeze-out, photodissociation, various heating-cooling mechanisms and chemical reactions). 

This sample of transition disks with large cavities provides a unique opportunity for transition disk studies: not only do we have spatially resolved submillimeter observations of continuum and CO for all disks, also the stellar properties such as temperature, luminosity, mass and accretion rate are very well constrained by VLT X-shooter data \citep{Alcala2014, Alcala2017}. Gas and dust masses and upper limits were derived for the full sample \citep{Ansdell2016,Miotello2017}, and for 36 of the detected disks the dust surface density profiles were analyzed using a two-layer disk model \citep{Tazzari2017}. The properties of transition disks derived here can thus be compared directly with primordial disks within a star forming region. As the sample of transition disks with large cavities is taken from a complete disk survey, general properties and statistics can be derived directly and compared with exoplanet statistics of giant planets at wide orbital radii. Disk survey studies have suggested that there are different populations of transition disks with a different origin \citep{OwenClarke2012,Garufi2017}, which can be properly tested with the Lupus disk survey.  

The paper is structured as follows. In Sect. 2 we describe the details of the ALMA observations and the sample selection. The modeling approach is presented in Sect. 3. Section 4 presents the modeling results. Section 5 discusses the implications for planet formation. 

\section{Observations}
\subsection{Data reduction}
The observations were obtained during ALMA Cycle 2 in June 2015 in Band 7. The full details of the setup and calibration process are described in \citet{Ansdell2016}. The data were reduced and imaged using CASA 4.4.0. The images have a typical spatial resolution of 0.34"$\times$0.28", using Briggs weighting with a robust of -1. The spectral cubes of $^{13}$CO 3--2 and C$^{18}$O 3--2 were extracted when detected, with a spectral resolution of 1 km s$^{-1}$ and a typical rms of 24 mJy bm$^{-1}$ km s$^{-1}$. The S/N in the zero-moment maps reaches $\sim$15 for the brightest disks. The continuum images (335 GHz or 890 $\mu$m) have a typical rms of 0.25 mJy bm$^{-1}$ (M stars) and 0.4 mJy bm$^{-1}$ (K stars), resulting in a peak S/N of 50-100 on the continuum. As the observations were only 1-2 minutes per sources, the S/N is lower than in previous studies of gas in transition disks \citep[e.g.][]{vanderMarel2016-isot}.

\subsection{Sample}
The 11 transition disks of our sample (Table \ref{tbl:sample}) are selected from a complete disk survey of the Lupus star forming region (Lupus I-IV clouds) for stars with $M_*>$0.1 $M_{\odot}$ \citep[][Ansdell et al.\, in prep.]{Ansdell2016}. The total disk population contains 96 objects, as defined in Table 1 of \citet{Ansdell2016}, with the addition of Sz~91, Sz~76, Sz~77, Sz~102, V1094~Sco, EX~Lup, GQ~Lup and RXJ~1556.1-3655 that were not observed as part of the initial ALMA program, and with the omission of J16104536-3854547 and J16121120-3832197, which were found to be non-members in a study of the stellar properties \citep{Alcala2017}. Sz~91 was observed in a separate program (see below), and the other 7 additional sources were observed in similar settings in a recent ALMA Cycle 4 program, but none of these were found to have a dust cavity (Ansdell et al.\ in prep., van Terwisga et al.\ in prep.). For the remainder of this paper, we restrict our analysis to the transition disks with cavities $>$20 AU. The 11 targets in this study this form a \emph{full census of transition disks with large cavities ($>$20 AU) in a complete disk survey of a young star forming region.} 

Our sample (top part of Table \ref{tbl:tdstatus}) consists of all transition disks in Lupus I-IV as defined in \citet{Ansdell2016}, with the exception of  J16011549-4152351 and J16081497-3857145. These targets are omitted because there are no signs of a cavity in the continuum visibility curve (or image) implying that any cavity, if present, is smaller than $\sim$20 AU radius. We note that Sz~103 and Sz~104, not identified as transition disk in \citet{Ansdell2016}, have been identified in the literature as a transition disk candidate \citep{vanderMarel2016-spitzer}, but there are no signs of a cavity in the continuum visibility curve (or image) either and thus excluded from our analysis sample. Also Sz~76, part of the additional Cycle 4 program, was identified in this study as transition disk candidate based on the SED, but no cavity was resolved in the ALMA data. The SEDs of the five unresolved transition disk candidates from Table \ref{tbl:tdstatus} analyzed by \citet{vanderMarel2016-spitzer} all indicate a cavity size of 2-5 AU, which indeed cannot be resolved with our spatial resolution. 

\citet{Tazzari2017} identified Sz~129 and J16000236-4222145 as potential transition disks based on their fit of a negative $\gamma$ in the surface density profile, but as there is no clear evidence for a cavity in the continuum visibility curve these targets are omitted from our sample. We note that Tazzari's findings are consistent with a shallow surface density in the inner part of the disk and a sharper outer radius cutoff, unlike the sharp inner cutoffs seen in our sample.

Finally, we have added Sz~118 and Sz~91 to the sample of transition disks to be analyzed. Sz~118 was not identified as transition disk candidate before, but the continuum image and visibility curve show signs of a dust cavity. Sz~91 was not included in the original Lupus sample \citep{Ansdell2016} and therefore not part of our dataset, but its transition disk status was confirmed through SED analysis and continuum imaging \citep{Canovas2015,Canovas2016,vanderMarel2016-spitzer}. We use unpublished ALMA archival data from program 2012.1.00761.S (PI Tsukagoshi) of the Band 7 continuum and the $^{12}$CO 3--2 line to constrain the properties of this disk. These data were reduced using the provided pipeline reduction script with CASA version 4.3.1. These data were imaged using natural weighting, resulting in a beam size of 0.18$\times$0.15". The continuum rms is 55 $\mu$Jy beam$^{-1}$ and the line rms is 5 mJy beam$^{-1}$ per 0.5 km s$^{-1}$ window.

The sample thus consists of 10 new transition disks in spatially resolved submillimeter observations, out of which 6 had been identified before through their SED based  on photometric points. Only Sz~91 was previously confirmed as a transition disk in submillimeter interferometry \citep{Canovas2015}. Table \ref{tbl:tdstatus} furthermore lists for the line data whether the cavity is spatially resolved.

The targets are located in the LupIII cloud at distances between 150 and 200 pc (see Table \ref{tbl:sample}), except for MY~Lup which is in the LupIV cloud. The stellar properties of the targets in the sample are taken from \citep{Alcala2017}, where the stellar mass and accretion rate estimates are derived from evolutionary models \citep{Siess2000}. These properties are listed in Table \ref{tbl:sample}. J16070854-3914075 is classified as a flat source according to its infrared properties and the stellar properties are highly uncertain, as discussed in detail in \citet{Alcala2017}. However, assuming their spectral type of M5 with an effective temperature of 3125 K, we find that their derived stellar luminosity of 0.011 $L_{\odot}$ with $A_v$=3.6 mag is too low to reproduce both the stellar photosphere and the thermal dust emission of the disk. Instead, we use $L_* = 0.18 L_{\odot}$, which was derived using a method for a protostar with an envelope that reprocesses part of the stellar radiation \citep{Evans2009} and is more consistent with the properties of YSOs in Lupus \citep{Alcala2017}. We stress that its derived disk properties remain highly uncertain due to the unconstrained stellar properties.

\begin{table*}[!ht]
\small
\begin{center}
\caption{Sample}
\label{tbl:sample}
\begin{tabular}{lllllllllll}
\hline
\hline
Target&RA&Dec&SpT&T$_{\rm eff}$&$L_*$&$M_*$&$\log \dot{M}$&$d$\\
&&&&(K)&($L_{\odot}$)&($M_{\odot}$)&($M_{\odot}$ yr$^{-1}$)&(pc)\\
\hline
Sz~91&16 07 11.576&-39 03 47.88&M1&3705&0.31&0.47&-8.7&200\\
J16083070-3828268&16 08 30.688&-38 28 27.27&K2&4900&3.0&1.8&-9.1&200\\
Sz~111&16 08 54.672&-39 37 43.50&M1&3705&0.33&0.46&-9.1&200\\
RY~Lup&15 59 28.373&-40 21 51.63&K2&4900&1.7&1.5&-8.2&150\\ 
Sz~118&16 09 48.641&-39 11 17.29&K5&4350&1.1&1.1&-9.0&200\\
Sz~123A&16 10 51.577&-38 53 14.18&M1&3705&0.20&0.46&-8.8&200\\
Sz~84&15 58 02.505&-37 36 03.10&M5&3125&0.12&0.18&-9.3&150\\
Sz~100&16 08 25.750&-39 06 01.69&M5.5&3057&0.17&0.18&-9.5&200\\
J16102955-3922144&16 10 29.542&-39 22 14.89&M4.5&3200&0.16&0.22&-9.8&200\\
J16070854-3914075$^a$&16 07 08.539&-39 14 07.94&M5&3125&0.18&0.17&-9.2&200\\
MY~Lup&16 00 44.503&-41 55 31.33&K0&5100&0.78&1.0&$<$-9.7&150\\
\hline
\end{tabular}
\end{center}
$^a$ The stellar properties of J16070854-3914075 remain highly uncertain as this is a flat infrared source \citep{Alcala2017}.
\end{table*}

\begin{table*}[!ht]
\small
\begin{center}
\caption{Transition disk status}
\label{tbl:tdstatus}
\begin{tabular}{llllll}
\hline
\hline
Target&Identification&Continuum\footnote{The mark "resolved" refers to whether the cavity is resolved in the image.}&$^{13}$CO&C$^{18}$O&$F_{\rm 890{\mu}m}$ (mJy)\\
\hline
Sz~91&\citet{Canovas2015,Canovas2016}&resolved&resolved\footnote{$^{12}$CO 3--2 rather than $^{13}$CO 3--2 line}&-&62$\pm$0.1\\
J16083070&\citet{Merin2008}, vdM2016b&resolved&resolved&resolved&135$\pm$1.1\\
Sz~111&\citet{Merin2008},vdM2016b&resolved&resolved&resolved&179$\pm$1.0\\
RY~Lup&Not previously identified&resolved&resolved&marg. resolved&276$\pm$1.2\\
Sz~118&Not previously identified&resolved&unresolved&non-detection&63$\pm$1.0\\
Sz~123A&\citet{Merin2008}&marg. resolved&unresolved&non-detection&41$\pm$0.6\\
Sz~84&\citet{Merin2010},vdM2016b&marg. resolved&unresolved&non-detection&33$\pm$0.4\\
Sz~100&Not previously identified&marg. resolved&unresolved&non-detection&55$\pm$0.6\\
J16102955&\citet{Bustamante2015},vdM2016b&marg. resolved&unresolved&non-detection&7.1$\pm$0.35\\
J16070854&Not previously identified&marg. resolved&unresolved&non-detection&92$\pm$1.5\\ 
MY~Lup&vdM2016b&marg. resolved&unresolved&non-detection&177$\pm$0.8\\
\hline
Sz~76&vdM2016b&no sign cavity&unresolved&non-detection&9.2$\pm$0.2\\
Sz~103&vdM2016b&no sign cavity&non-detection&non-detection&11.5$\pm$0.2\\
Sz~104&\citet{Merin2008}, vdM2016b&no sign cavity&non-detection&non-detection&3.2$\pm$0.2\\
Sz~112&vdM2016b&no sign cavity&non-detection&non-detection&3.9$\pm$0.2\\
J16011549&vdM2016b&no sign cavity&resolved&resolved&82$\pm$0.9\\
J16081497&\citet{Bustamante2015}&no sign cavity&non-detection&non-detection&8.3$\pm$0.3\\
Sz~129&\citet{Tazzari2017}&no sign cavity&unresolved&non-detection&181$\pm$0.5\\
J16000236&\citet{Tazzari2017}&no sign cavity&resolved&non-detection&120$\pm$0.6\\
\hline
\end{tabular}
\end{center}
\end{table*}

\section{Modeling}
In order to constrain the size and depth of dust and gas cavities, we use the physical-chemical modeling code DALI \citep{Bruderer2012,Bruderer2013}. Our modeling approach is similar to that in \citet{vanderMarel2016-isot}. As physical-chemical modeling is computationally expensive, MCMC or $\chi^2$ fitting is not practical. The combined dust and gas data provide sufficient constraints for a  model of the gas and dust surface density profile consistent with the data. However, we stress that the derived model parameters are not necessarily unique.

The position angle (east of north) and inclination are derived with the continuum images and the $^{13}$CO spectrum and given in Table \ref{tbl:fitresults}. They are consistent within errors with the estimates found by \citet{Ansdell2016} and \citet{Tazzari2017}. Errors on these values are typically 5$^{\circ}$.

\subsection{Physical model}
\begin{figure}
\begin{center}
\includegraphics[width=0.5\textwidth]{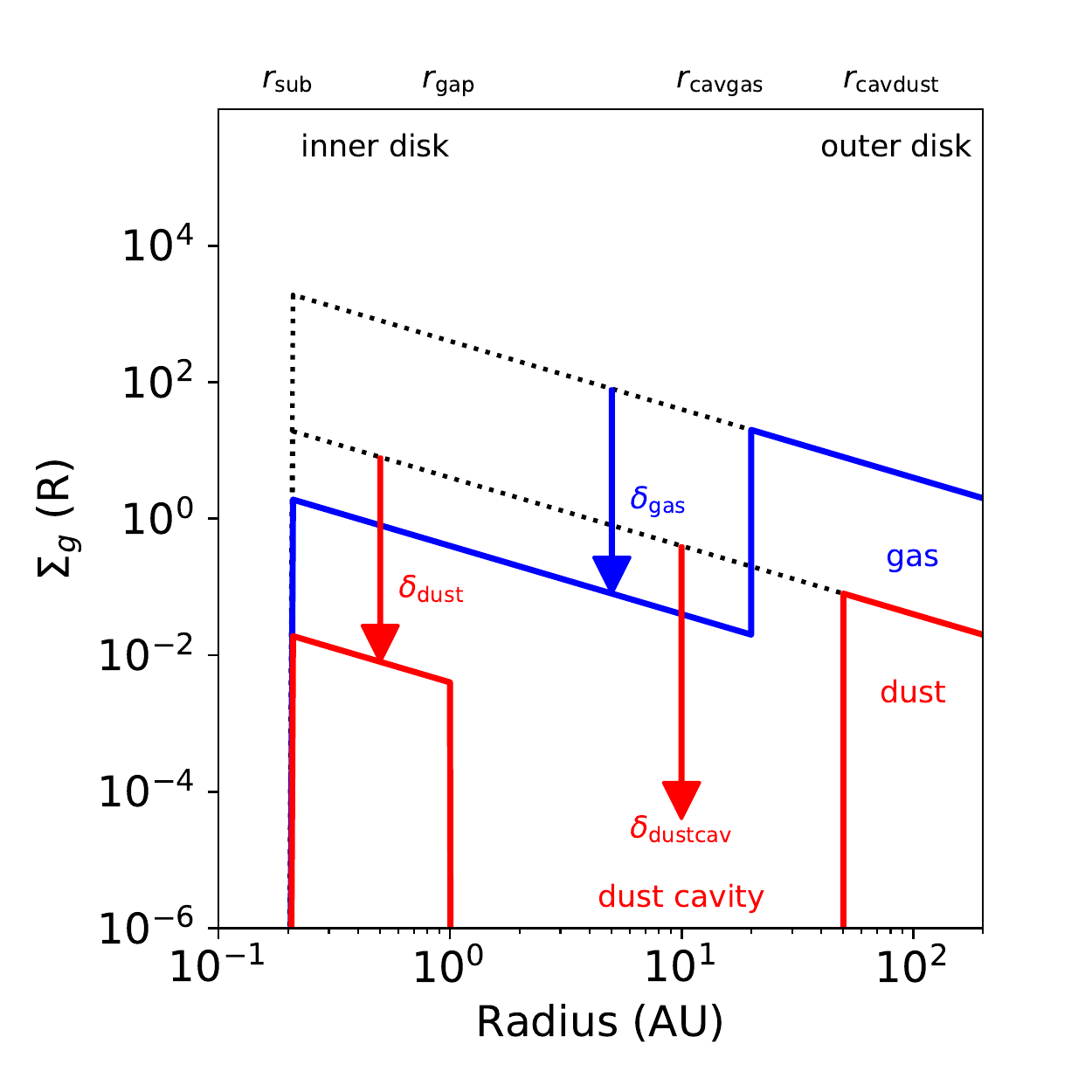}
\caption{Generic disk surface density model. Note that $r_{\rm gap}$ is fixed to 1 AU in our models.}
\end{center}
\label{fig:genericmodel}
\end{figure}

As a starting point for our models we adopted the physical structure suggested by \citet{Andrews2011}, as implemented by \citet{Bruderer2013} and fully described in \citet{vanderMarel2015-12co}. The surface density $\Sigma(r)$ is assumed to be a radial power-law with an exponential cut-off following the time-dependent viscosity disk model $\nu \sim r^{\gamma}$ with $\gamma=1$  \citep{LyndenPringle1974,Hartmann1998}
\begin{equation}
\Sigma(r) = \Sigma_c \left(\frac{r}{r_c}\right)^{-\gamma} {\rm exp}\left(-\left(\frac{r}{r_c}\right)^{2-\gamma}\right)
\end{equation}
The gas and dust follow the same density profile, but the gas-to-dust ratio is varied throughout the disk, as shown in Figure \ref{fig:genericmodel}. Inside the cavity, the dust density is zero, except for the inner disk, which is set by $\delta_{\rm dust}$. The inner disk ranges between the sublimation radius $r_{\rm sub}$ (AU)=0.07$\sqrt{L_* {\rm (L_{\odot})}}$ and $r_{\rm gap}$=1 AU. The gas density inside the cavity is varied with $\delta_{\rm gas}$. For disks where the gas cavity in the CO zero-moment map is resolved, the gas cavity radius is defined separately as $r_{\rm cavgas}$, while in the other disks $r_{\rm cavgas}=r_{\rm cavdust}$. In the outer disk, the gas-to-dust ratio is set to GDR=100. The outer radius of the model grid is set to 400 AU, although the detectable emission and thus the disk outer radius is usually smaller. As the outer radius is not the main goal of the analysis, this is not further explored.   

The vertical structure is defined by the scale height $h_c$ and the flaring angle $\psi$, following $h(r)=h_c(r/r_c)^{\psi}$.  The fraction of large grains $f_{\rm ls}$ and the scale height of the large grains $\chi$ are set to 0.85 and 0.2 respectively to describe the settling. These parameters are kept fixed as settling does not significantly affect the derived structure of gas and dust in the disk \citep{Bruderer2013}. Stellar properties and accretion rates are taken from \citet{Alcala2014,Alcala2017} and listed in Table \ref{tbl:sample}. More details on the implementation of the stellar radiation fields (including UV excess), the dust composition, settling, and vertical structure are given in \citet{vanderMarel2015-12co,vanderMarel2016-isot}. The DALI module for isotope-selective photodissociation \citep{Miotello2014} is not included to limit the model computational times. This may overpredict the computed C$^{18}$O fluxes.

\newpage
\subsection{Modeling approach}
For all targets, the following step-by-step procedure is used to find a model that is consistent with all data using manual fitting, similar as used in \citet{Bruderer2014,vanderMarel2015-12co}. We stress that this model may not be a unique solution but provides constraints on the cavity size in the dust and the amount of gas inside the cavity, the key parameters of interest. Degeneracies between disk parameters in transition disk modeling have been discussed in detail in the literature \citep[e.g.][and references therein]{Mathews2012}.

\begin{enumerate}
\item Find first estimates of $\Sigma_c$ and $r_{\rm cav}$ roughly to the continuum visibility curve, in particular the total flux and the location of the null.
\item Adjust $h_c$ and $\psi$ to fit the infrared excess in the SED.
\item Adjust $\delta_{\rm dust}$ to fit the near infrared excess (inner disk) in the SED.
\item Adjust $r_c$ and $r_{\rm cav}$ (keeping the total dust mass the same with $\Sigma_c$) to fit the continuum visibility curve (Figure \ref{fig:dustmodel}).
\item Find the highest possible value for $\delta_{\rm dustcav}$ (drop of dust density inside the cavity) that is still consistent with the dust continuum visibility curve and image.
\item Set the gas-to-dust ratio to 100 in the outer disk, set $\delta_{\rm gas}$=1 (no drop in gas density inside the cavity), compute the $^{13}$CO emission and compare with the data.
\item Vary the $r_c$ (and corresponding $h_c$ and $\psi$ parameters) if necessary to fit the $^{13}$CO emission in the outer disk: as $^{13}$CO emission is partially optically thick, the temperature structure is relevant. Variations in $h_c$ and $\psi$ are typically less than 0.05 at this stage.
\item Vary $\delta_{\rm gas}$ and $r_{\rm cavgas}$ (for the disks where the $^{13}$CO emission is spatially resolved) to fit the inner part of the $^{13}$CO zero-moment map and visibility curve.
\end{enumerate}

The best fit parameters are given in Table \ref{tbl:fitresults}. The C$^{18}$O data have too low S/N to constrain the structure, but the integrated fluxes are compared with the model output. Only dust masses are given as these are directly fit to the continuum flux.

\section{Results}
\subsection{Final models}
\begin{figure*}[!ht]
\includegraphics[width=\textwidth, trim=150 100 150 150]{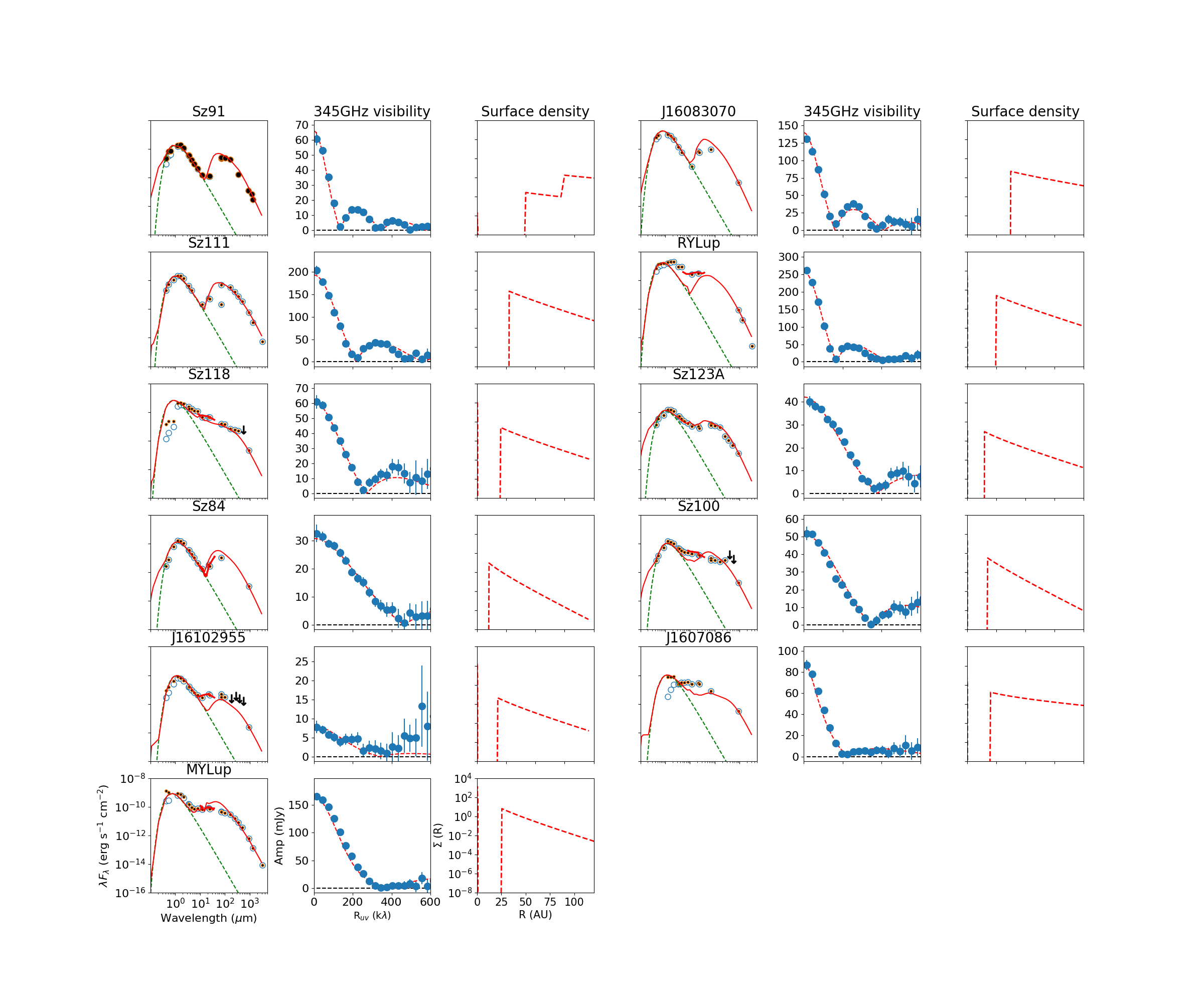}
\caption{Results dust modeling: Spectral Energy Distribution, continuum visibility curve, and density plot. The dereddened data points in the SED shown in black dots and the observed data in blue circles. In the visibility curves, data are given by blue dots with error bars. The model is overplotted on both panels in red. }
\label{fig:dustmodel}
\end{figure*}

\begin{figure}[!ht]
\begin{center}
\includegraphics[width=0.7\textwidth]{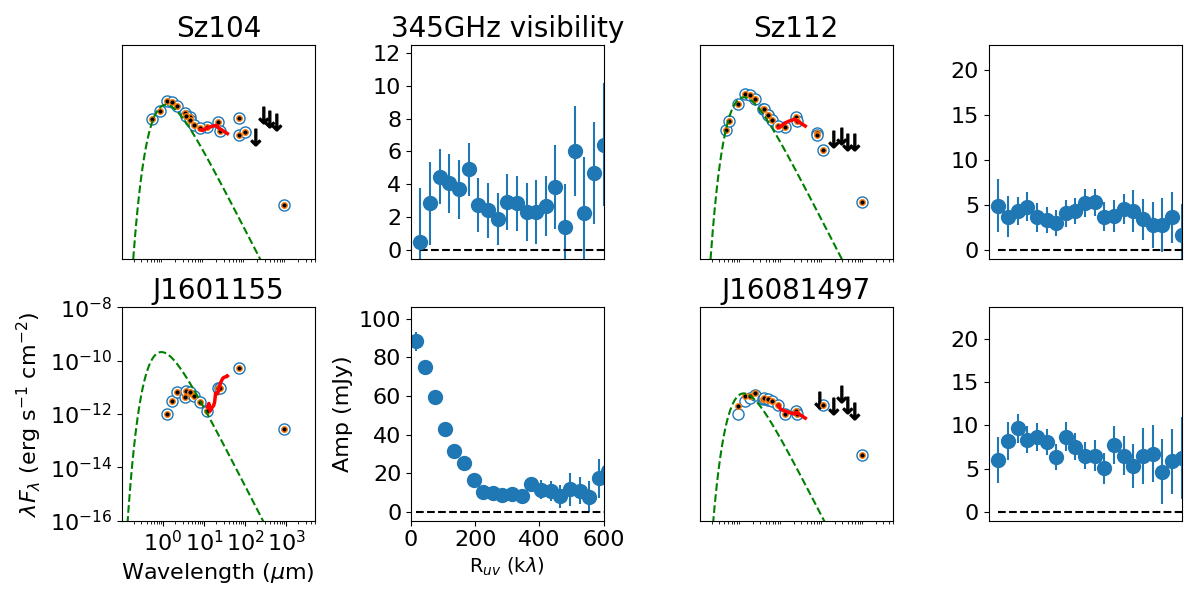}
\caption{SEDs and visibility plots of the transition disk candidates for which the cavity could not be confirmed.}
\end{center}
\label{fig:nocavity}
\end{figure}

\begin{figure*}[!ht]
\includegraphics[width=\textwidth, trim=150 100 150 150]{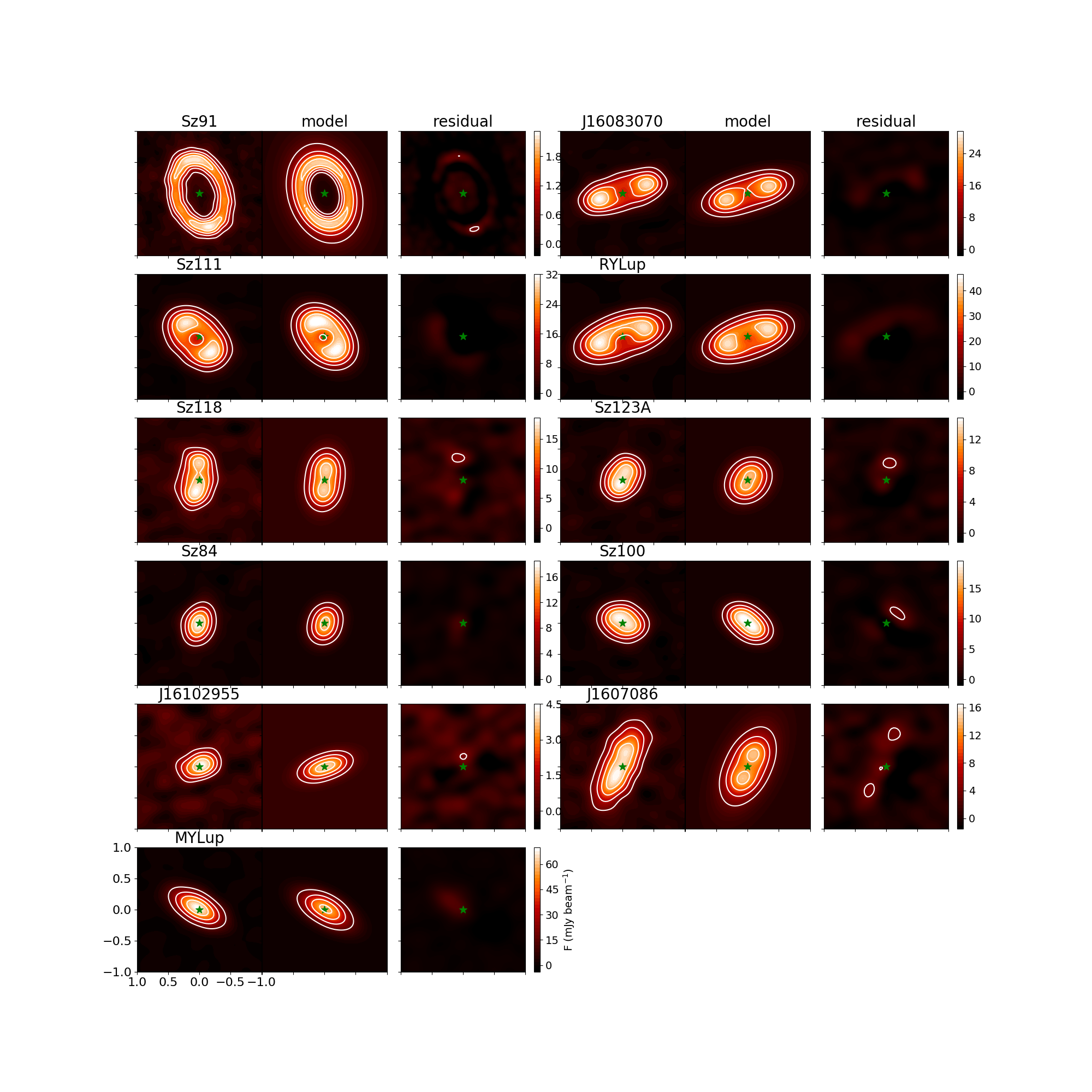}
\caption{Results of dust modeling. Best fit for the continuum image, continuum model and residual. The contours indicate the 20, 40, 60 and 80\% levels of the maximum flux density in the image.}
\label{fig:bestdustmodel}
\end{figure*}

\begin{figure*}[!ht]
\includegraphics[width=\textwidth, trim=150 100 150 150]{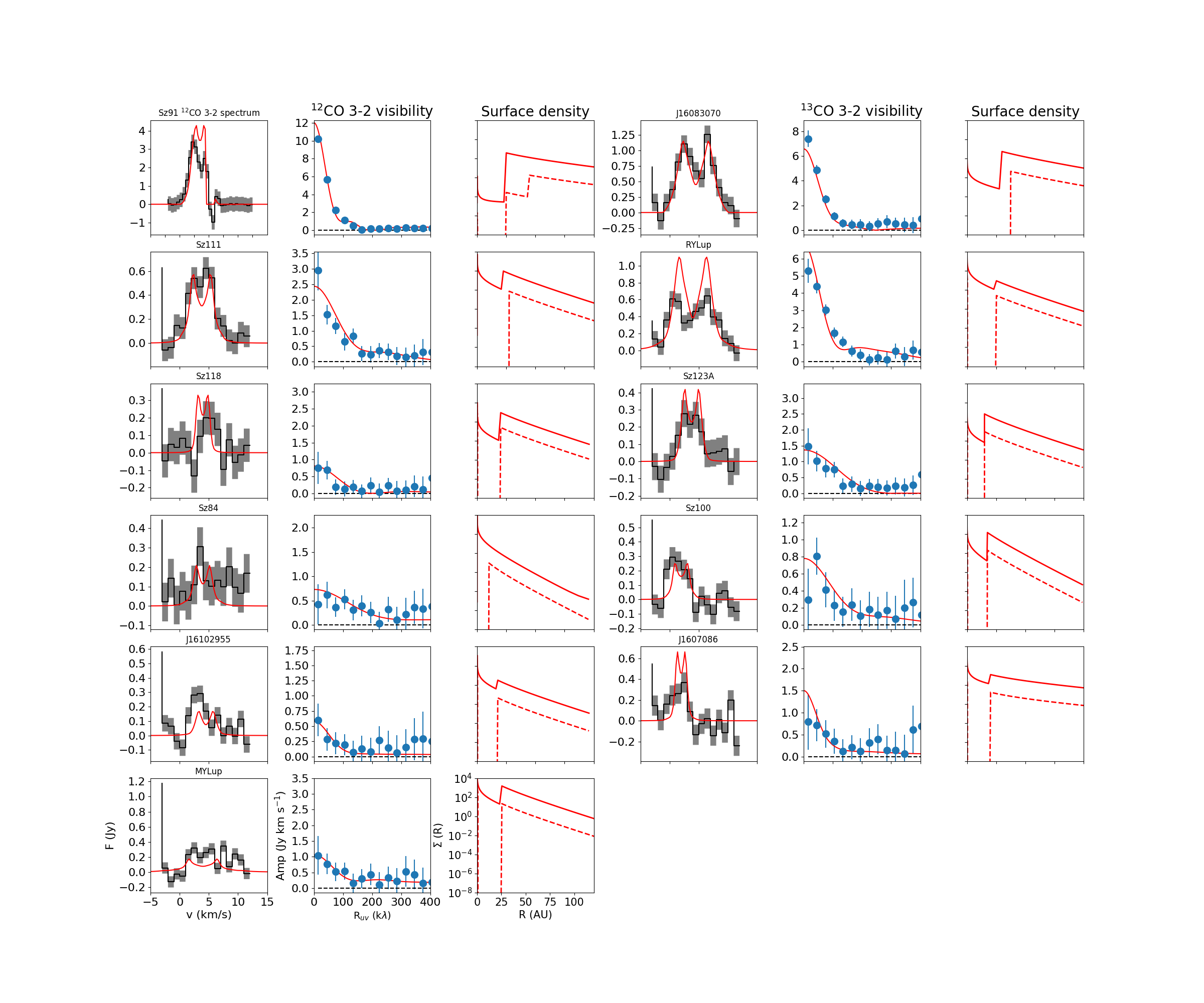}
\caption{Results of gas modeling. From left to right: $^{13}$CO 3--2 spectrum, visibility curve of the integrated $^{13}$CO 3--2 and density plot. The data are shown in black with grey noise levels and the best-fit model in red. In the density plot, the solid line indicates the gas and the dashed line the dust.} 
\label{fig:gasmodel}
\end{figure*}

\begin{figure*}[!ht]
\includegraphics[width=\textwidth, trim=150 100 150 150]{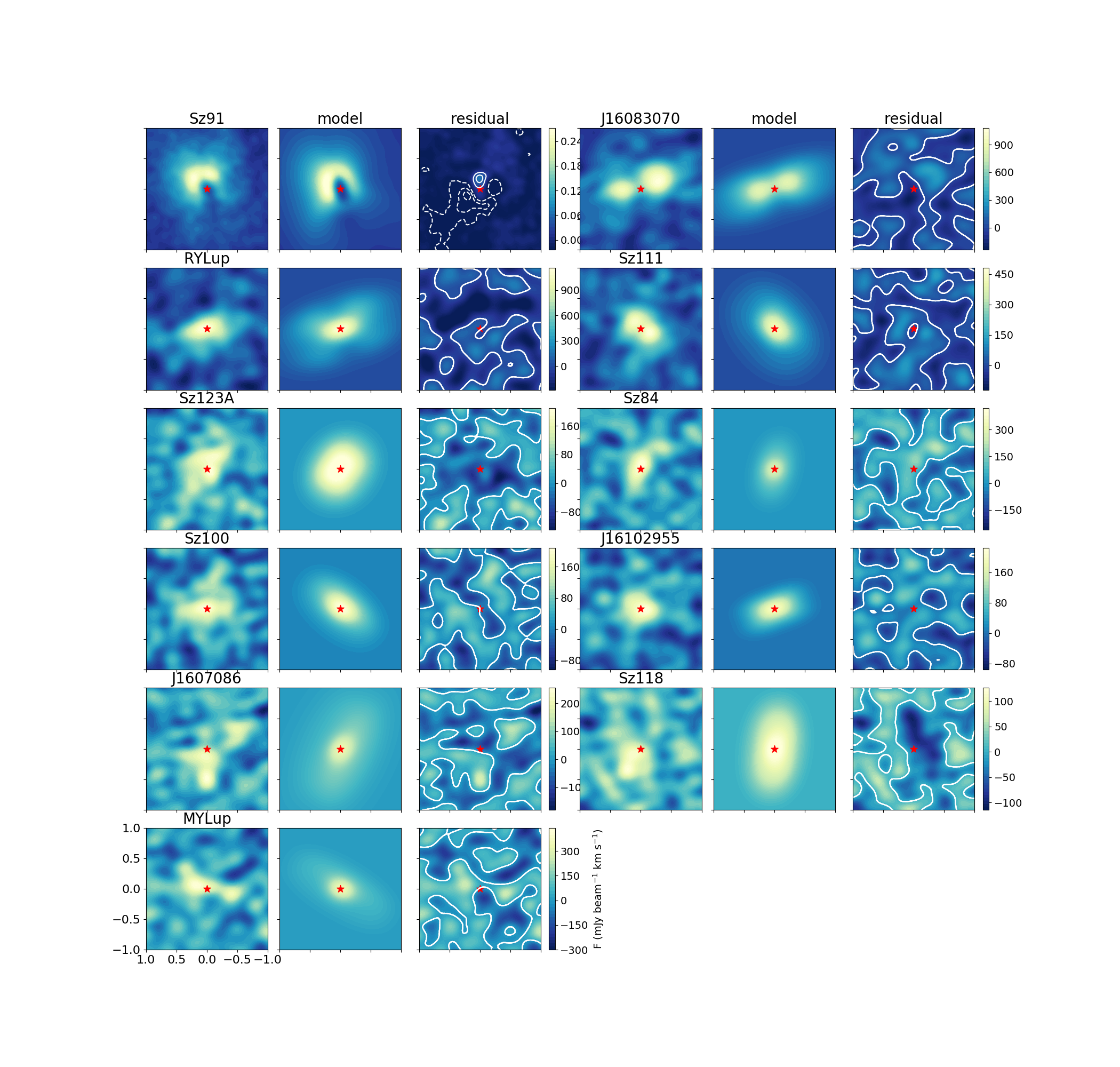}
\caption{Results of gas modeling. Best fit for the $^{13}$CO 3--2 zero-moment map image, model and residual for each target. For Sz~91 the $^{12}$CO 3--2 moment map is shown; in the model image, the velocities affected by foreground absorption are masked out.} 
\label{fig:bestgasmodel}
\end{figure*}

Figure \ref{fig:dustmodel} to \ref{fig:bestgasmodel} show the results of our modeling procedure and Table \ref{tbl:fitresults} provides the values of the fit parameters. For the visibility curves the amplitudes rather than real parts are shown, as several of the disk images appear non-axisymmetric due to optical depth effects by their high inclination. In this approach, all flux is included in the comparison. 

The final models generally agree well with the data visibility curves. Residuals in the images in Figure \ref{fig:bestdustmodel} and \ref{fig:bestgasmodel} likely originate from the image reconstruction of the visibilities, as the visibilities itself match well with the models, in particular the location of the null and the total flux.

The C$^{18}$O images are not shown due to the low S/N, but Figure \ref{fig:gasmodel} shows the spectra and integrated visibilities compared with the models for the $^{13}$CO line. The integrated fluxes of the C$^{18}$O 3--2 are given in Table \ref{tbl:emissionc18o}. The model fluxes are about a factor $>$2 higher than the fluxes of the data, indicating that the real gas-to-dust ratio of these disks may be $<$100, similar to previous modeling results of the Lupus disk masses \citep{Ansdell2016,Miotello2017}. Alternatively, the model fluxes of C$^{18}$O may be overestimated as the isotope-selective photodissociation  \citep{Miotello2014} was not included in this work. \citet{Miotello2014} showed that neglecting isotope-selective photodissociation can result in an underestimate of the disk mass. As the total disk mass is not the main focus of this work, the gas mass is not further explored. This implies that the $\delta_{\rm gas}$ values are uncertain by a factor of a few due to the uncertainty in $\Sigma_c$, which is lower than the typical uncertainty on $\delta_{\rm gas}$ itself, which is about a factor 10 (see Section \ref{sct:deltagas}).

The derived dust masses are within a factor of 10 of the derived dust masses compared to previous estimates where full disk models were assumed \citep{Ansdell2016,Miotello2017,Tazzari2017}. \citet{Ansdell2016} converted the integrated fluxes into dust masses for all the disks with a monochromatic temperature to obtain a first order dust mass distribution; \citet{Miotello2017} derived disk masses using the integrated fluxes and a large grid of DALI models; \citet{Tazzari2017} modelled the radial structure of the continuum emission of each individual source, fitting the visibilities with the two-layer disk model from \citet{ChiangGoldreich1997}, and did not include highly-inclined disks or disks with obvious dust cavities (several of our targets) in their sample. 

Since these past works assumed full disks without dust cavities, their estimates in comparison with the data are less reliable for the sample of transition disks that they analyze. In particular, dust masses of highly-inclined disks may be underestimated \citep{Miotello2017}. Furthermore, the inclusion of a dust cavity can significantly decrease the required dust mass for the same continuum flux. Our work focuses in particular on the radial (and vertical) structure of each individual source, using a full radiative transfer model, the SED and spatially resolved continuum data, rather than on the disk masses, so the differences in the derived disk mass are acceptable. 

Uncertainties in the derived model parameters remain due to redundancies, the noise levels, optical depth effects and assumptions in the physical-chemical modeling \citep[e.g.][]{Bruderer2013}. Typical uncertainties are $\sim$5 AU in the cavity radii and about one order of magnitude for the gas surface density (gas to dust ratio in the outer disk and $\delta_{\rm gas}$).

\begin{table*}[!ht]
\begin{center}
\caption{Modeling results}
\label{tbl:fitresults}
\begin{tabular}{l|lll|ccll|ll|ll}
\hline
\small
Target&\multicolumn{3}{c|}{Surface density}&\multicolumn{4}{c|}{Radial structure}&\multicolumn{2}{c|}{Vertical}&\multicolumn{2}{c}{Orientation}\\
\hline
&$r_c$&$\Sigma_{c,dust}$&$M_{\rm dust}$&$r_{\rm cavdust}$&$r_{\rm cavgas}$&$\delta_{\rm gas}$&$\delta_{\rm dust}$&$h_c$&$\psi$&PA&$i$\\
&(AU)&(g cm$^{-2}$)&($M_{\rm Earth}$)&(AU)&(AU)&&&&&($^\circ$)&($^\circ$)\\
\hline
Sz~91&	75&	0.07&	25&	90&	50&$\leq$	10$^{-5}$&	10$^{-20}$&	0.1&	0.3&	17&	45\\
J16083070&	50&	0.39&	38&	75&	60&	$\leq$10$^{-4}$&	10$^{-20}$&	0.10&	0.10&	107&	74\\
Sz~111&		25&	15&	232&	55&	45&	$\leq$10$^{-2}$&	10$^{-20}$&	0.08&	0.10&	40&	53\\
RY~Lup&		25&	2.0&	76&\multicolumn{2}{c}{50}&	$\geq$10$^{-1}$&	10$^{-1}$&	0.15&	0.15&	109&	68\\
Sz~118&		25&	1.0&	57&\multicolumn{2}{c}{40}&$\leq$10$^{-3}$&	10$^{-1}$&	0.07&	0.1&	173&	65\\
Sz~123A&	25&	0.36&	16&\multicolumn{2}{c}{30}&$\leq$10$^{-3}$&	10$^{-2}$&	0.15&	0.15&	145&	43\\
Sz~84&		15&	0.50&	6.4&\multicolumn{2}{c}{20}&-&	10$^{-20}$&	0.10&	0.05&	168&	65\\
Sz~100&		15&	8&	43&\multicolumn{2}{c}{35}&-&	10$^{-1}$&	0.1&	0.3&	50&	65\\
J16102955&	25&	0.25&	16&\multicolumn{2}{c}{35}&	-&	10$^{0}$&	0.15&	0.10&	110&	82\\
J16070854&	100&	0.10&	146&\multicolumn{2}{c}{40}&	-&	10$^{-1}$&	0.10&	0.10&	155&	65\\
MY~Lup&		15&	60&	616&\multicolumn{2}{c}{25}&-&	10$^{-1}$&0.05&	0.30&	59&	73\\
\hline
\end{tabular}
\\
\footnotesize{Error bars on $\Sigma_c$ and the $\delta$ parameters are typically an order of magnitude, on the vertical structure parameters $\leq$0.05, on the cavity radii $\sim$5 AU and $\sim5^{\circ}$ on the orientation. See discussion in \citet{vanderMarel2016-isot}.}
\end{center}
\end{table*}

\begin{table}
\begin{center}
\caption{Integrated C$^{18}$O 3--2 fluxes as observed and final models}
\label{tbl:emissionc18o}
\begin{tabular}{llll}
\hline
Target&Flux data&Flux final model\\
&(Jy km s$^{-1}$)&(Jy km s$^{-1}$)\\
\hline
J16083070&1.7$\pm$0.2&2.9\\
RY~Lup&1.1$\pm$0.2&2.6\\
Sz~111&0.70$\pm$0.07&1.0\\
Sz~123A&$<$0.10&0.44\\
Sz~84&$<$0.12&0.14\\
Sz~100&$<$0.10&0.30\\
J16102955&$<$0.10&0.23\\
J16070854&$<$0.18&0.34\\
Sz~118&$<$0.23&0.22\\
MY~Lup&$<$0.27&0.41\\
\hline
\end{tabular}
\end{center}
\end{table}

\begin{figure*}[!ht]
\includegraphics[width=\textwidth, trim=150 100 150 150]{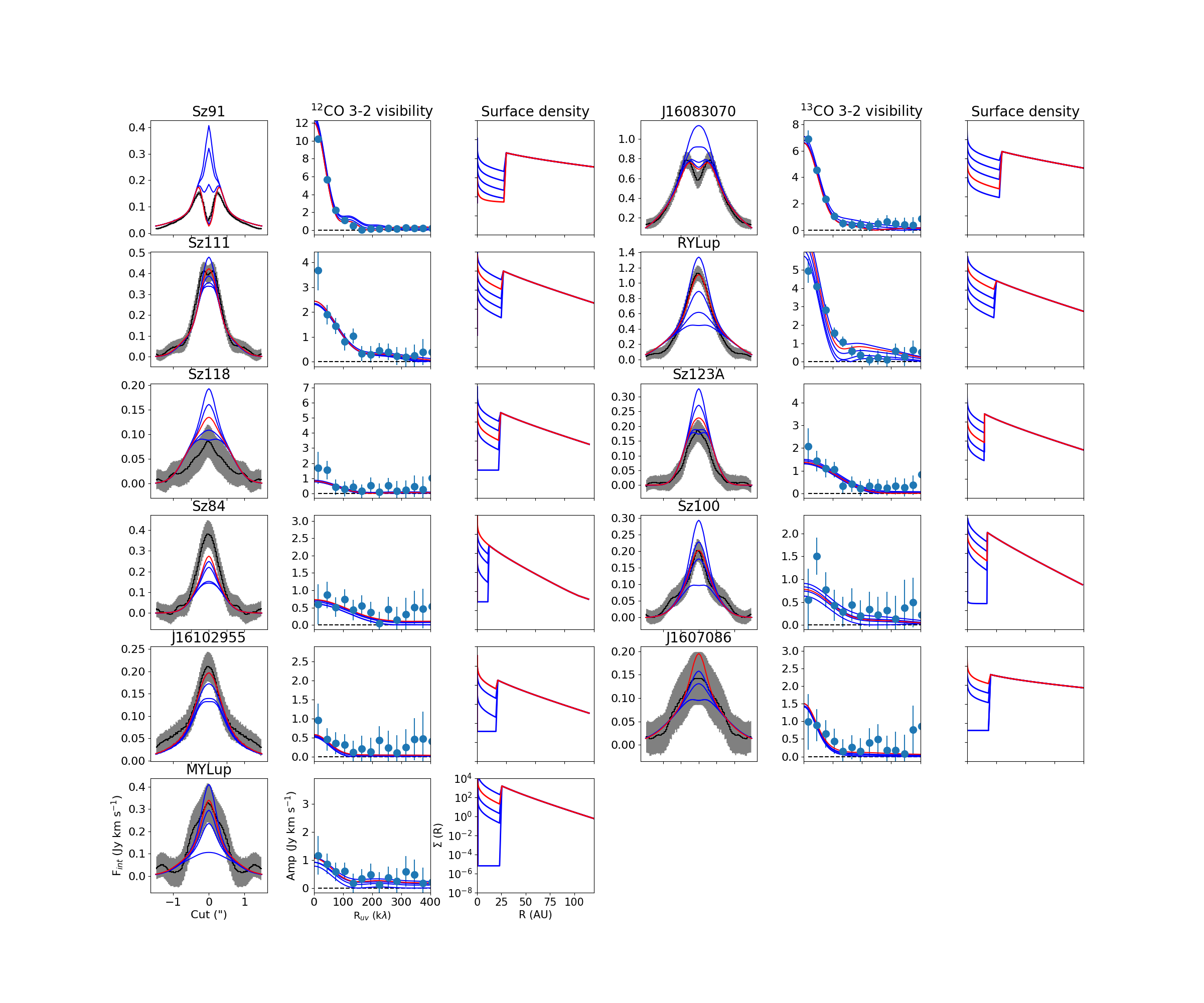}
\caption{Modeling results for the final gas model for different amounts of gas inside the gas cavity ($\delta_{\rm gas}$). From left to right, for each target, are shown: the azimuthally averaged cut of the integrated moment map of the $^{13}$CO 3--2; the integrated $^{13}$CO 3--2 visibility curve; the density profiles. The data are given in black, the different models in red (best-fit) and blue, for $\delta_{\rm gas}$=10$^{-1}$,10$^{-2}$,10$^{-3}$,10$^{-4}$,10$^{-5}$ and 10$^{-20}$. Note that not all models are run for the unresolved cases and there is no significant difference there between the different $\delta_{\rm gas}$ values. The plots demonstrate that the depth of the gas cavity is at least two orders of magnitude for half of the targets, but could not be quantified for the other disks.}
\label{fig:deltagas}
\end{figure*}

\subsection{Quantification of gas and dust inside the cavity}
\label{sct:deltagas}
In order to quantify the maximum amount of gas and dust inside the cavity, for each model we vary both $\delta_{\rm dustcav}$ and $\delta_{\rm gas}$. Neither of these parameter could be fit to a single value in any of these disks: they are merely upper limits. The results are shown in Figure \ref{fig:deltagas} and \ref{fig:deltadust}. Note that the gas models are run with $\delta_{\rm dustcav}$ set to zero (empty cavity).

Figure \ref{fig:deltadust} demonstrates that the depth of the dust cavity is at least two orders of magnitude, depending on the target. The well-resolved cavities (Sz~91, J16083070, Sz~111 and RY~Lup) are even better constrained, with a depth of at least three orders of magnitude. Fitting the dust continuum of Sz~91 requires a small amount of dust between 50 and 90 AU.

The amount of gas inside the cavities is more difficult to constrain. For the resolved cases (Sz~91, J16083070 and Sz~111) the gas cavity radius is smaller than the dust cavity radius and the gas gaps are deep, with $\delta_{\rm gas}\leq10^{-2}$. This is similar to previous results of gas cavities that have been analyzed with the same methodology \citep{vanderMarel2016-isot,Dong2017}. The other disks are analyzed using $r_{\rm cavgas}=r_{\rm cavdust}$ with different values of $\delta_{\rm gas}$, but the depth could only be constrained for two sources (Sz~118 and Sz~123A) to be $\delta_{\rm gas}\leq10^{-2}$. If the gas cavity is in reality smaller, it is expected to be even deeper. 

RY~Lup, although it has a large dust cavity, does not show any evidence of a gas gap in the $^{13}$CO, and quantifying the amount of gas actually shows $\delta_{\rm gas}\geq10^{-1}$, which is different from the other disks. The visibility curve of its $^{13}$CO data suggests that the gas cavity may be smaller, considering the location of the null compared with the models, but this remains difficult to constrain without the cavity seen in the image. On the other hand, there is potentially a warp in this disk (see Section \ref{sct:rylup}), which makes the quantification of the gas through the $^{13}$CO less reliable. For the other five disks it was not possible to set any constraints on $\delta_{\rm gas}$.

\clearpage

\section{Discussion}
\subsection{Clearing mechanism}
\label{sct:origin}
This transition disk study reveals dust cavities with radii of $\geq$20 AU for 11 disks, out of 96 disks studied in the Lupus star forming region. This is the first census of transition disks with large cavities ($>$20 AU) in a star forming region based on a complete disk survey. Four of these transition disks (RY~Lup, Sz~118, Sz~100 and J16070854) show a regular SED without a clear indication of a cavity as an infrared deficit, which would not have been recognized from photometric SED analysis alone. The IRS spectrum at 5-35 $\mu$m of RY~Lup \citep{Kessler2006} reveals a mid-infrared dip in the underlying continuum, but as the spectrum is dominated by the strong silicate features this is not seen in photometry, underlining the importance of the mid infrared spectroscopy.

\begin{figure}[!ht]
\begin{center}
\includegraphics[width=0.5\textwidth]{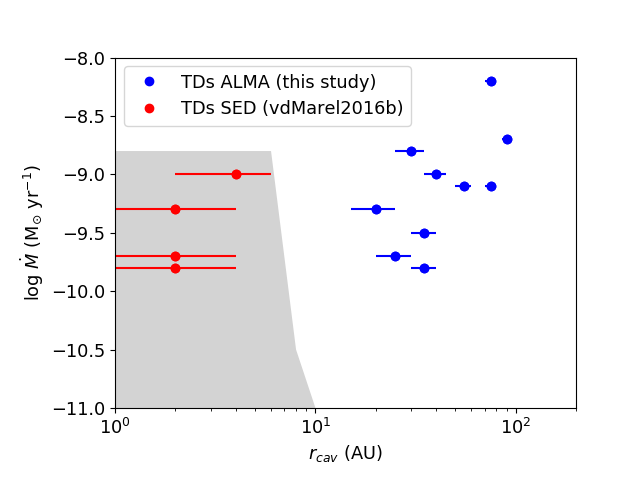}
\caption{Trend between accretion rates and derived dust cavity radii for the resolved ALMA disks in this study (blue) and the cavity sizes from SED modeling from \citet{vanderMarel2016-spitzer} (red). The grey area indicates the regime where cavities can be explained by photoevaporation models, following \citet{Owen2011,Owen2011erratum}. The transition disks with large cavities from this study fall outside this regime.}
\label{fig:photoevap}
\end{center}
\end{figure}

Common explanations for the origin of transition disk dust cavities include photoevaporation \citep[e.g.][]{Alexander2014}, dead zone instabilities \citep[e.g.][]{Regaly2012} and clearing by a substellar or planetary companion \citep{LinPapaloizou1979}. All disks have a measured accretion rate \citep{Alcala2017}, which makes, in combination with the large cavity radius the photoevaporation mechanism an unlikely explanation, as demonstrated in e.g. \citet{Owen2011,Ercolano2017} and Figure \ref{fig:photoevap}. Figure \ref{fig:photoevap} also shows the properties of the disks that remained unresolved in this study but for which the SED analysis indicates a small dust cavity radius \citep{vanderMarel2016-spitzer}. These disks are listed at the bottom of Table \ref{tbl:tdstatus} and fall within the photoevaporation regime. On the other hand, \citet{Ercolano2017xray} suggest that cavities in disks with modest gas depletion of carbon and oxygen could be explained by X-ray photoevaporation out to cavity radii as large as 100 AU with accretion rates of $\sim10^{-8}$ M$_{\odot}$ yr$^{-1}$. We note that the two disks with large cavities for which the accretion rate is compatible with chromospheric emission (MY~Lup and J16070854) have cavity sizes of 25 and 40 AU respectively, which still put them outside the range of transition disks that can be reproduced by photoevaporation models \citep{Ercolano2017}. 

A dead zone and planet gap will both create a (radial) dust trap at its outer edge, which gives the appearance of a dust ring \citep{Pinilla2012b}. The key to distinguish between these two mechanisms is the distribution of the gas: a dead zone will only show a minor change in gas surface density, whereas a companion can lower the gas surface density significantly up to several orders of magnitude, depending on its mass and the disk viscosity \citep[e.g.][]{Zhu2011,Fung2014,Pinilla2016dz}. 

In our sample of 11 transition disks, it was possible to constrain the gas surface density inside the cavity for about half of the targets. In 3 cases, the gas cavity is found to be smaller than the dust cavity, similar to previous high resolution studies \citep{Bruderer2014,Zhang2014,vanderMarel2016-isot,Dong2017,Boehler2017}. This is consistent with planet-disk interaction models in combination with dust trapping, as the dust trap is expected to be at the outer edge of the gas gap and the gas cavity radius thus further in than the dust cavity radius. Note that our approximation of the surface density profile with sharp drops is not physically realistic, and the gas likely follows a smooth drop such as seen in J1604-2130 \citep{Dong2017}. Such a profile cannot be constrained from our observations, but it was found to be consistent with data for several transition disks in \citet{vanderMarel2016-isot}. For the other disks, the gas profile inside the cavity remains unresolved and only the amount of gas inside the dust cavity was varied through $\delta_{\rm gas}$. 

For half of the targets, the gas cavities have a density drop of several orders of magnitude, which is consistent with clearing by a massive companion (several Jupiter masses) in combination with low values of viscosity ($\alpha\sim10^{-4}$), according to comparisons with planet-disk interaction models \citep{Fung2014}. A more detailed discussion on typical companion masses based on $\delta_{\rm gas}$ values is given in \citet{vanderMarel2016-isot}. The only exception is RY~Lup, which does not have a deep gas gap. For the other disks, a deep gas cavity cannot be ruled out or confirmed with the available resolution. It is striking that apart from RY~Lup there are no counter examples of resolved gas cavities with the same radius as their dust cavities in both this study and the literature. This appears to be a common phenomenon in transition disks. This suggests that transition disks with large cavities are caused by giant planets of several Jupiter masses at several tens of AU. The case of RY~Lup is further discussed in section \ref{sct:rylup}.

Whether the gaps inferred from $^{13}$CO observations are really several orders of magnitude deep remains a topic of discussion. \citet{Facchini2017gaps} notice that the gas temperature can drop due to heating-cooling effects of the dust grains when dust evolution effects are taken into account. In our modeling, dust evolution is not explicitly taken into account, but the depletion of dust in the cavity mimics the effect. However, it is possible that the gas gaps may be less deep (by an order of magnitude) than inferred here due to additional heating-cooling effects by the dust. The derived gas cavity radius is unlikely to be affected by this effect as the gas temperature would decrease inside the entire dust-depleted region, whereas the $^{13}$CO emission shows a smaller cavity radius than the dust emission.

\subsection{Gas cavity vs gap}
The derived gas surface density close to the star can be compared with the expected gas surface density based on the accretion rate onto the star, using Eqn. 5 from \citet{Manara2014} and the measured accretion rates \citep{Alcala2017}. Assuming a steady-state viscous disk with $\alpha$=10$^{-2}$ and our derived values for $\Sigma(r)$, the gas surface density at 1 AU based on our models and from the accretion rates are given in Table \ref{tbl:acc}.

\begin{table}[!ht]
\begin{center}
\caption{Accretion properties}
\label{tbl:acc}
\begin{tabular}{lll}
\hline
Target&$\Sigma_{\rm 1 AU}$&$\Sigma_{\dot{M}_{\rm acc}}$\\
&(g cm$^{-2}$)&(g cm$^{-2}$)\\
\hline
Sz~91&$<$0.005&22\\ 
 J16083070&$<$0.19&17\\ 
 Sz~111&$<$360&8.6\\ 
 RY~Lup&$>$480&124\\ 
 Sz~118&$<$2.4&17\\ 
 Sz~123A&$<$0.86&17\\ 
 Sz~84$^a$&$<$702&3.4\\ 
 Sz~100$^a$&$<$11226&2.1\\ 
 J1610295$^a$&$<$600&1.2\\ 
 J1607085$^a$&$<$990&-\\ 
 MY~Lup$^a$&$<$84195&$<$5.1\\
\hline
\end{tabular}
\\
$^a$ $\delta_{\rm gas}$ completely unconstrained: value for $\Sigma_{\rm 1 AU}$ is derived assuming $\delta_{\rm gas}$=1.\\
\end{center}
\end{table}

\begin{figure*}[!ht]
\includegraphics[width=\textwidth]{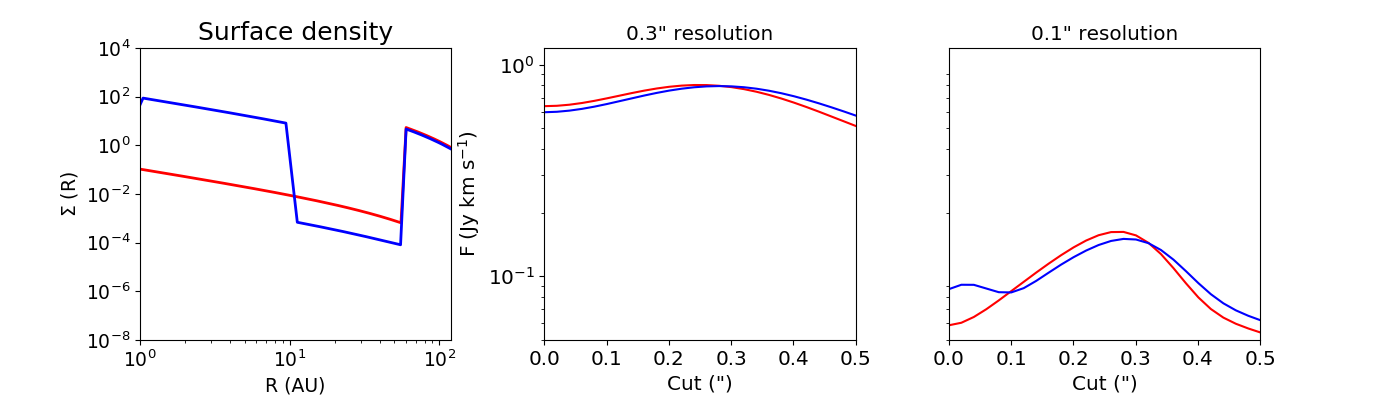}
\caption{Radial cuts of a gas cavity and a gas gap for the $^{13}$CO 3--2 emission of the J16083070 model at 0.3" and 0.1" spatial resolution, with the modeled emission in colors and the data in black (left panel only). The surface density profile is given in the left panel. At the resolution of the observations (middle), these two profiles cannot be distinguished, which means that the origin of the cavity cannot be properly constrained.}
\label{fig:gasgap}
\end{figure*}

For the five disks for which $\delta_{\rm gas}$ remained unconstrained the limits for $\Sigma_{\rm 1 AU}$ are much higher than $\Sigma_{\dot{M}_{\rm acc}}$, suggesting that $\delta_{\rm gas}$ is in fact lower than 1. The values for $\Sigma_{\rm 1 AU}$ for the disks where $\delta_{\rm gas}$ is constrained are generally lower than based on the measured accretion rates \citep{Alcala2017}, with the exception of RY~Lup (discussed above) and Sz~111, although the derived value is an upper limit. This suggests that the gas surface density close to the star is in reality higher than measured from the CO observations \citep{Najita2015}, and the surface density profile is actually a gap rather than a cavity, with an inner gas disk inside the cavity. In our modeling setup this inner gas disk is not included to limit the number of free parameters. Figure \ref{fig:gasgap} demonstrates that the spatial resolution of $\sim$0.3" is insufficient to distinguish between a full cavity and a gap at the measured gap sizes of a few tens of AU for the case of J160803070, the disk with the largest gas cavity in our sample. In a disk with a gap, the inner part of the disk needs to be continuously replenished with material flowing through the gap \citep[e.g.][]{Rosenfeld2014,WangGoodman2017}. A gap is more consistent with the predictions of planet-disk interaction models of the effect of a planet on a disk \citep{Fung2014}, whereas a cavity is consistent with for example a dead zone in combination with a MHD wind, a scenario suggested by \citet{Pinilla2016dz}, so higher resolution observations are required in order to properly constrain the structure of the gas. 

\subsection{Transition disks and spectral types}
\begin{table}[!ht]
\begin{center}
\caption{All known transition disks from millimeter imaging}
\label{tbl:alltds}
\begin{tabular}{llll|llll}
\hline
Name&SpT&Origin&Ref$^1$&Name&SpT&Origin&Ref$^1$\\
\hline
V892~Tau		&B8	&	Tau	&	1, 18	&
HD~169142	&	B9	&	Isolated	&	2, 19	\\
HD~34282	&	B10	&	Isolated	&	3, 20	&
AB~Aur	&	A0	&	Tau/Aur	&	4, 21	\\
IRS~48	&	A0	&	Oph	&	5, 22	&
HD~97048	&	A0	&	Isolated	&	6, 23	\\
MWC~758	&	A8	&	Isolated	&	7, 24	&
HD~100453	&	A9	&	Isolated	&	8, 18	\\
HD~135344B	&	F4	&	Isolated	&	9, 25	&
HD~142527	&	F6	&	Isolated	&	9, 26	\\
RY~Tau	&	G1	&	Tau	&	10, 27	&
SR~21	&	G3	&	Oph	&	11, 25	\\
LkH$\alpha$~330	&	G3	&	Tau	&	11, 25	&
T~Cha	&	K0	&	Cham	&	9, 28	\\
MY~Lup	&	K0	&	Lup	&	12, 29	&
RY~Lup	&	K2	&	Lup	&	12, 29	\\
J16073080	&	K2	&	Lup	&	12, 29	&
SR~24S	&	K2	&	Oph	&	13, 24	\\
CS~Cha	&	K2	&	Cham	&	11, 30	&
SZ~Cha	&	K2	&	Cham	&	11, 30	\\
UX~TauA	&	K2	&	Tau	&	14, 24	&
LkCa~15	&	K3	&	Tau	&	11, 24	\\
DoAr~44	&	K3	&	Oph	&	11, 24	&
J1604-2130	&	K5	&	UppSco	&	15, 31	\\
GM~Aur	&	K5	&	Tau	&	11, 24	&
PDS~70	&	K5	&	Isolated	&	16, 18	\\
Sz~118	&	K5	&	Lup	&	12, 29	&
RXJ~1615-3255	&	K7	&	Isolated	&	11, 24	\\
Sz~91	&	M0	&	Lup	&	12, 32	&
RXJ~1633-2429	&	M0	&	Oph	&	15, 33	\\
Sz~111	&	M1	&	Lup	&	12, 29	&
DM~Tau	&	M1	&	Tau	&	11, 24	\\
Sz~123A	&	M1	&	Lup	&	12, 29	&
MHO~2	&	M3	&	Tau	&	17, 18	\\
J16102955	&	M4	&	Lup	&	12, 29	&
Sz~84	&	M5	&	Lup	&	12, 29	\\
J160708054	&	M5	&	Lup	&	12, 29	&
Sz~100	&	M5.5	&	Lup	&	12, 29	\\
WSB~60	&	M6	&	Oph	&	11, 24	\\

\hline
\end{tabular}
\\$^1$First reference regards the spectral type, the second the (sub)millimeter image.
\\Refs: 1. \citet{Wahhaj2010}, 2. \citet{Dunkin1997}, 3. \citet{Merin2004}, 4. \citet{Hernandez2004}, 5. \citet{Brown2012a}, 6. \citet{Irvine1977}, 7. \citet{Chapillon2008}, 8. \citet{Vieira2003}, 9. \citet{Schisano2009}, 10. \citet{Cieza2010}, 11. \citet{Manara2014}, 12. \citet{Alcala2017}, 13. \citet{Spezzi2008}, 14. \citet{GarciaLopez2006}, 15. \citet{Lawson2004}, 16. \citet{Riaud2006}, 17. \citet{Luhman2000}, 18) ALMA archive,
19) \citet{Fedele2017} ,
20) \citet{vanderPlas2017},
21) \citet{Pietu2005},
22) \citet{vanderMarel2013},
23) \citet{vanderPlas2016},
24) \citet{Andrews2011},
25) \citet{Brown2009},
26) \citet{Casassus2013},
27) \citet{Isella2010rytau},
28) \citet{Huelamo2015},
29) This work,
30) \citet{Pascucci2016},
31) \citet{Mathews2012},
32) \citet{Canovas2015},
33) \citet{Cieza2012}	
\end{center}
\end{table}

\begin{figure}[!ht]
\begin{center}
\includegraphics[width=0.5\textwidth]{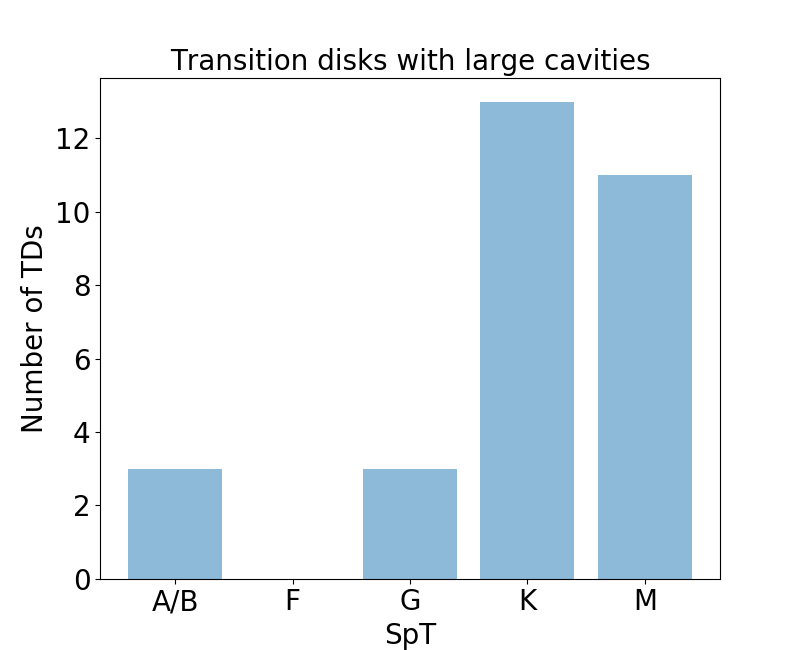}
\caption{The stellar distribution of all known transition disks in Lupus, Chamaeleon, Taurus and Ophiuchus.}
\label{fig:imf}
\end{center}
\end{figure}

The transition disks in this study extend the sample of confirmed transition disks with large cavities based on submillimeter imaging down to late spectral types. In Table \ref{tbl:alltds} we present an overview of all known transition disks from millimeter imaging with their spectral type and origin. Sample studies of transition disks in submillimeter imaging are biased towards early type stars, as these disks are brighter. Furthermore, these studies include a large number of isolated Herbig stars. Although isolated Herbigs are rare, the sample is likely to be complete within a distance of 200 pc due to their brightness, whereas later type isolated stars are much harder to detect. However, we exclude the isolated objects from Table \ref{tbl:alltds} in order to construct a stellar distribution of the young (1-3 Myr) star forming regions Lupus, Chamaeleon, Taurus and Ophiuchus  (Figure \ref{fig:imf}). We exclude Upper Sco as it is known to be significantly older \citep{Carpenter2014}. The Lupus transition disks are expected to be complete down to cavity radii of 20 AU (this work) and for Chamaeleon I a disk survey at lower spatial resolution was presented in \citet{Pascucci2016}. Although full ALMA disk surveys of Taurus and Ophiuchus are not yet available, these regions have been well studied with the SubMillimeter Array, and we expect that the majority of transition disks with large cavities are covered. If any transition disks are missing in this plot, they are expected to be found around K and M stars, as all early type stars in these regions have been targeted in millimeter interferometric studies. Figure \ref{fig:imf} reveals that the known transition disks span a full spectral range with a peak at the late-type stars, similar to the general spectral type distribution in star forming regions \citep[e.g.][]{Luhman2012}, suggesting that transition disks can be found amongst all spectral classes.

\subsection{Comparison with exoplanet statistics}
\label{sct:exoplanets}

\begin{figure*}[!ht]
\includegraphics[width=\textwidth]{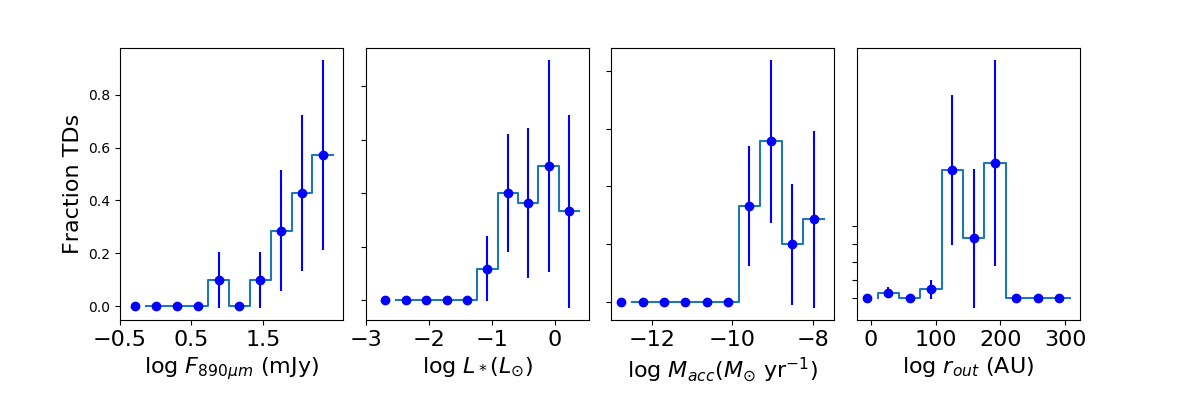}
\caption{Histogram of the fraction of transition disks in Lupus as a function of their properties, as a function of millimeter-flux $F_{\rm 890{\mu}m}$, stellar luminosity $L_*$, accretion rate $M_{\rm acc}$ and dust outer radius $R_{\rm out}$ (measured using a curve of growth method on the continuum images, taking the outer radius at 90\% of the flux).}
\label{fig:cmlplot}
\end{figure*}

The fraction of transition disks with large cavities in Lupus is $\gtrsim$11\% ($\pm$3.7\%), which has important consequences for the interpretation of the cavity origin. Previous estimates of transition disk fraction range between 5 and 25\%, depending on the definition, selection criteria and sample \citep[e.g.][and references therein]{OwenClarke2012,vanderMarel2016-spitzer}. Previous estimates were often based on studies where transition disks were selected based on either SED analysis or millimeter imaging. Selection based on millimeter imaging was biased towards the brightest objects available due to the limited sensitivity of pre-ALMA submillimeter interferometers. On the other hand, SED-based selection includes transition disks with smaller cavities of a few AU (albeit with high uncertainty), which is likely incomplete due to the uncertainties in SED analysis, in particular for photometric SEDs without the IRS spectrum: more than one third of our targets would not have been identified as transition disk based on the SED alone. Besides, disks with small dust cavities fall within the photoevaporation regime \citep{Owen2011} rather than planet clearing and may thus be a different class of transition disks (Figure \ref{fig:photoevap}). 

It is unlikely that many other transition disks with cavity sizes $\geq$20 AU are missed in this study due to sensitivity: the faintest transition disk in this sample is J16102955-3922144 with a total flux of 7 mJy, which is in the lowest quartile of the detections, and even there a null at $\sim$400 k$\lambda$ was readily detected. 17 disks have a detected flux below 7 mJy, but all of these are unresolved. 27 disks in Lupus remain undetected with an upper limit of $\sim$1 mJy. The SEDs of the disks with fluxes $<$7 mJy do not show evidence for large cavities either. 

Figure \ref{fig:cmlplot} displays the distribution of the transition disks as function of observables millimeter-flux $F_{\rm 890{\mu}m}$  (corrected for distance), stellar luminosity $L_*$, accretion rate $M_{\rm acc}$ and dust outer radius $R_{\rm out}$. The latter is measured using a curve of growth method on the images, taking the outer radius at 90\% of the flux. Figure \ref{fig:cmlplot} demonstrates that transition disks with large cavities are millimeter-bright disks with large outer radii, as suggested by \citet{OwenClarke2012}. They distinguish two populations of transition disks in a combined study of SED and millimeter imaging targets: millimeter-faint disks with low accretion rates and small cavity sizes, and millimeter-bright disks with high accretion rates and large cavity sizes, suggesting that these two populations thus have different origins. As our study selects only disks with large cavities, we are likely seeing only the second population. Considering that our study is based on spatially resolved transition disks within a complete and deep flux-limited disk survey, our derived fraction of $\gtrsim$11\% is a more reliable estimator of the large cavity transition disks.  

Clearing by giant planets at wide orbits appears to be the most likely explanation for transition disks with large cavities, but the observed fraction poses an important issue. Planet detections in transition disks are rare, and often debated \citep[e.g.][]{KrausIreland2012,Quanz2013,Quanz2015,Reggiani2014,Biller2014,Sallum2015,Thalmann2016}. Exoplanet population studies around main sequence stars demonstrate that the occurrence rate of planets of 5-13 $M_{\rm Jup}$ at 10-300 AU is less than 8\% around all stars, and even lower for FGK ($<$6.8\%) and M ($<$4.2\%) stars \citep{Bowler2016}, which is below the observed transition disk fraction of $\gtrsim$11\%, although the derived percentages also depend on assumptions regarding the planet brightness and assumed planetary evolutionary model (hot-start vs cold-start).   

If the inconsistency is not caused by the biases in the exoplanet statistics, planet migration may be able to explain the lack of giant planets at wide orbits.
Giant planets are expected to migrate inwards rapidly due to planet-disk interaction \citep{KleyNelson2012} if the disk mass is comparable to or greater than the planet mass, but the high transition disk fraction and thus lifespan of the transition disk phase ($\sim$several 10$^5$ years) indicates that the migration time scale must be of similar length. Also, dynamical scattering \citep[the Lidov-Kozai mechanism, e.g.][]{WuMurray2003}, which happens at much longer time scales ($\sim$100 Myr), could explain the inconsistency between transition disk fraction and giant planet occurrence rate. Based on the low $\delta_{\rm gas}$ values, our analysis suggests that planets of Jupiter mass are responsible for the gas cavity, using the relations in \citet{Fung2014} for a low viscosity of $\alpha\sim10^{-4}$. Even lower viscosities could be explained by lower mass planets down to Super Earths, which would be undetected in current imaging surveys. It is also possible that rather than a single giant planet, multiple Neptune-like planets may be responsible for the large cavities.

\subsection{Transition disks vs primordial disks}
\begin{figure*}[!ht]
\includegraphics[width=\textwidth]{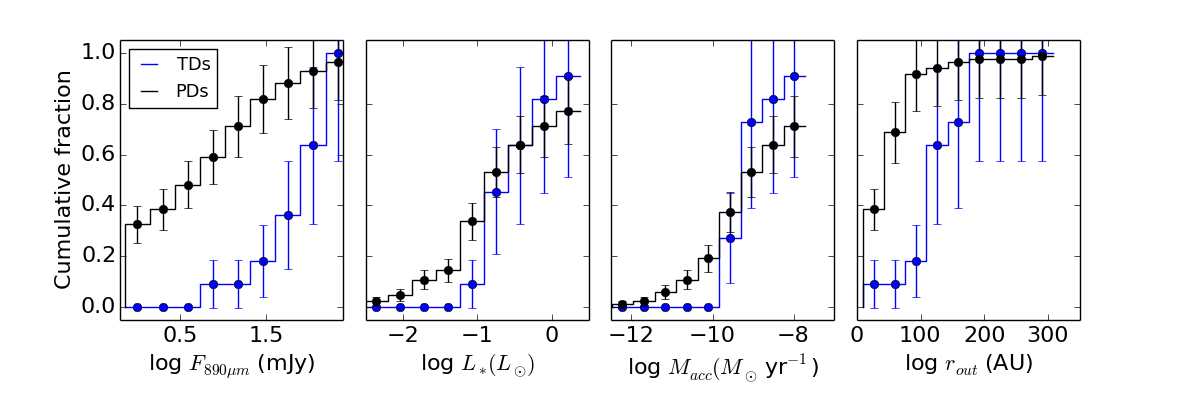}
\caption{Cumulative plot of both transition and primordial disk fractions in Lupus with respect to their total number in each bin, as a function of millimeter-flux $F_{\rm 890{\mu}m}$  (corrected for distance), stellar luminosity $L_*$, accretion rate $M_{\rm acc}$ and dust outer radius $R_{\rm out}$  (curve of growth: radius at 90\% of the flux).}
\label{fig:survcmlplot}
\end{figure*}

The transition disks as defined in this study obviously deviate from the primordial disks by the presence of a dust cavity. Any smaller dust cavities would not be resolved with the available resolution. One of the main questions in transition disk studies is whether all disks go through this 'transition phase', or whether only a subset will develop a cavity \citep{Cieza2007}. A remarkable property of the transition disks within Lupus is that they fall on the high tail of the dust mass distribution \citep{Ansdell2016}, as also seen in Taurus and Ophiuchus \citep{Andrews2013}. In Figure \ref{fig:survcmlplot} we compare the properties of transition disks with primordial disks using cumulative plots of the transition/primordial disk fraction as a function of observables: continuum flux, stellar luminosity, accretion rate and dust outer radius.

\subsubsection{Millimeter flux}
The transition disks occupy the brightest part of the millimeter flux distribution, whereas the primordial disks have a more even distribution of millimeter fluxes. 

\subsubsection{Stellar properties}
The stellar luminosity and mass accretion plots show no distinction in the distribution of primordial and transition disks with respect to $M_{\rm acc}$. \citet{Najita2015} and \citet{Manara2016} suggest that the transition disks in the $M_{\rm dust}-M_{\rm acc}$ fall below the main trend, indicating that they have either lower accretion rates or higher disk masses. Our plot suggests that the accretion rates are similar to primordial disks and the disk masses are higher.

\subsubsection{Dust outer radius}
The dust outer radius of transition disks with large cavities is clearly larger than that of primordial disks, but this is partially a selection effect as their cavity size sets the outer radius already at $\gtrsim$50  AU for a $\sim$20 AU cavity limit due to the spatial resolution. Still, more than 70\% of the transition disks have an outer radius of more than 120 AU, whereas 80\% of the primordial disks is smaller than 120 AU based on our $r_{\rm out}$ estimates. However, a handful of the primordial disks also have a significant extents of $>$120 AU: IM~Lup (Sz~82), Sz~83, Sz~98 and Sz~133. 

\begin{figure}[!ht]
\includegraphics[width=\textwidth]{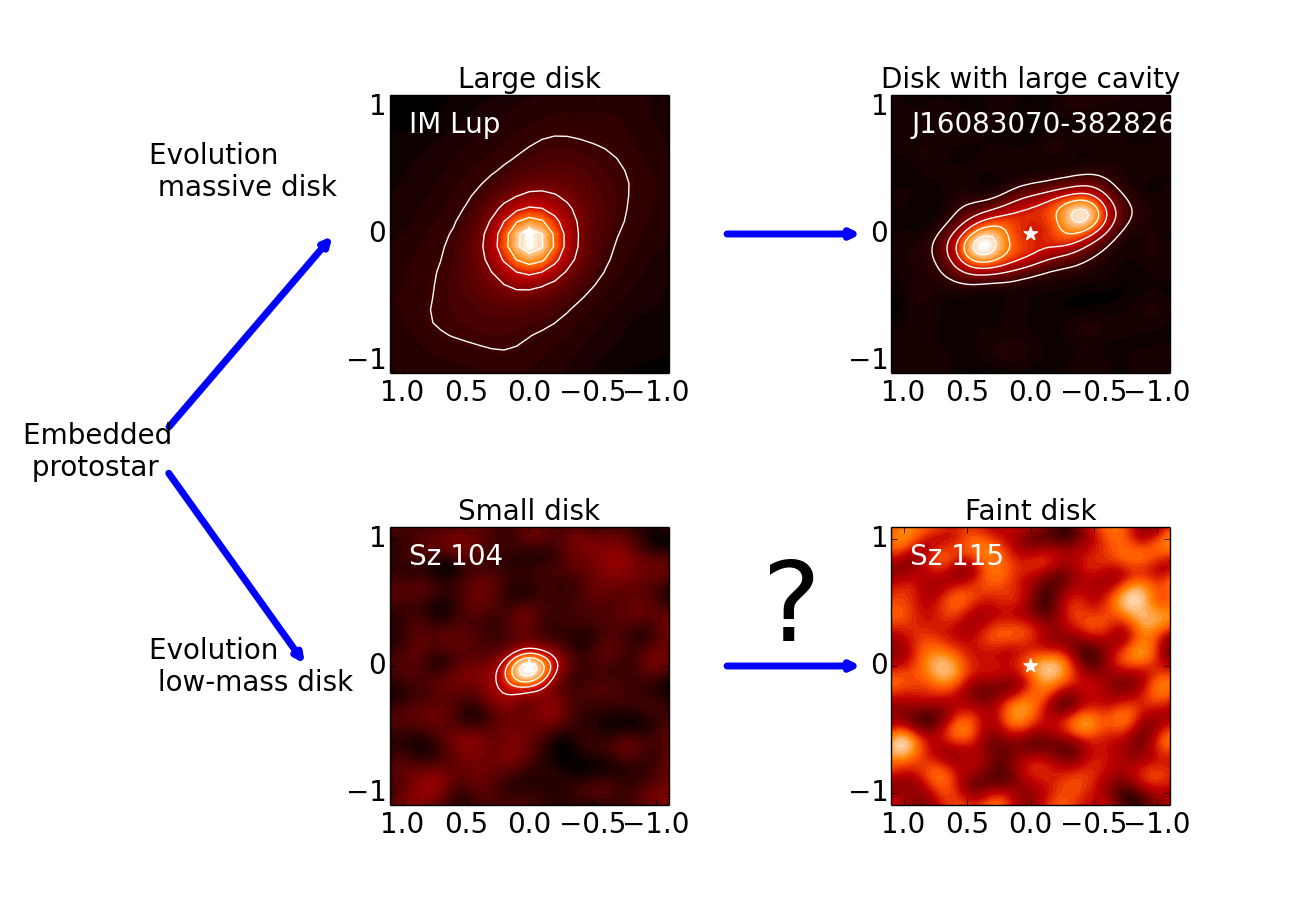}
\caption{Proposed evolutionary sequence for massive, extended disks towards transition disks with large cavities (top). The evolutionary sequence for low-mass disks with a small extent (bottom) remains unclear: either they evolve differently through general dissipation or as a scaled-down equivalent with smaller cavities and smaller extent. The sequence is illustrated using some typical examples from the Lupus disk survey.}
\label{fig:cartoon}
\end{figure}

\subsubsection{Two evolutionary pathways?}
In recent years, ALMA has revealed several primordial disks with significant extent in dust, usually with ring-like structures, such as HL~Tau \citep{HLTau2015} and HD163296 \citep{Isella2016}. Certain transition disks show evidence for both inner cavities and outer rings, such as HD~169142 \citep{Fedele2017}, HD~97048 \citep{vanderPlas2017}, HD~100546 \citep{Walsh2014} and RX~J1615-3255 \citep{vanderMarel2015-12co}. A large extent is inconsistent with our understanding of the role of radial drift \citep{Weidenschilling1977}, which acts to decrease the outer radius of the dust, such as observed in some disks \citep{Perez2012,Tazzari2016}. The large extent in both primordial and transition disks could be explained by the presence of unresolved dust rings in the outer disk as rings are the result of a mechanism that prevents the large dust grains from drifting in, similar to the edges of transition disk cavities. Possible mechanisms are e.g. giant planets creating dust traps at the edges or condensation fronts where dust grows more efficiently. We suggest that large, massive, ring-like primordial disks have an evolutionary connection with transition disks with large cavities, where a ring system evolves into a transition disk, as planet formation in the inner part is induced due to the radial concentrations of dust particles, resulting in (giant) planets clearing large cavities. Both the transition disks and the large primordial disks are on the high-mass end of the Lupus disk mass distribution, indicating that this evolutionary path occurs for more massive disks. The majority of Lupus disks are smaller and less massive and thus less likely to form giant planets, and it remains unclear if they evolve through a different pathway or are a scaled-down equivalent of this scenario (see Figure \ref{fig:cartoon}).

A division in evolutionary pathways has been suggested previously based on SED studies of \emph{Spitzer} data, where Class II objects were further divided in transition disk (mid infrared deficit or dip compared to the mean T Tauri star SED) or anemic disks (decrease of infrared excess at all wavelengths compared to the mean T Tauri star SED) \citep{Cieza2007,Currie2009}. It was suggested that disks either go through a gradual clearing process between Class II and III, or clearing from the inside out. Recently, \citet{Maaskant2013} and \citet{Garufi2017} suggested two different evolutionary pathways for Herbig stars, based on a reassessment of the properties of a large sample of Group I and Group II Herbig stars. Instead of the conventional evolution from flared Group I disks (high infrared excess) to Group II disks (low infrared excess) as a consequence of settling of grains, they propose that large Group II disks are in fact self-shadowed, and Group I disks gapped, where the gap is opened at a later stage. On the other hand, there are also small Group II disks which evolve into even smaller Group II disks. This scenario may be more broadly applicable to T Tauri stars with different spatial extents. 

Alternative to an actual division in evolutionary pathways, it is also possible that the second group of disks has smaller cavities and would thus be the scaled-down equivalents of the ring to cavity scenario. There are a handful of transition disks that have been identified through their SED with smaller cavities ($<$10 AU radius) that could not be resolved with the current ALMA data (Figure \ref{fig:nocavity} and Section 2). In general it is difficult to recognize small cavities from the SED, so the sample of transition disks with small cavities is potentially much larger. The high fraction of Mini-Neptunes and Super Earth planets found around main sequence stars at small orbital radii suggests that small gaps and cavities must be common. Higher resolution ALMA data of large samples of disks in star-forming regions are required to test this scenario.

\subsection{The peculiar case of RY~Lup}
\label{sct:rylup}
\begin{figure}[!ht]
\begin{center}
\includegraphics[width=0.5\textwidth]{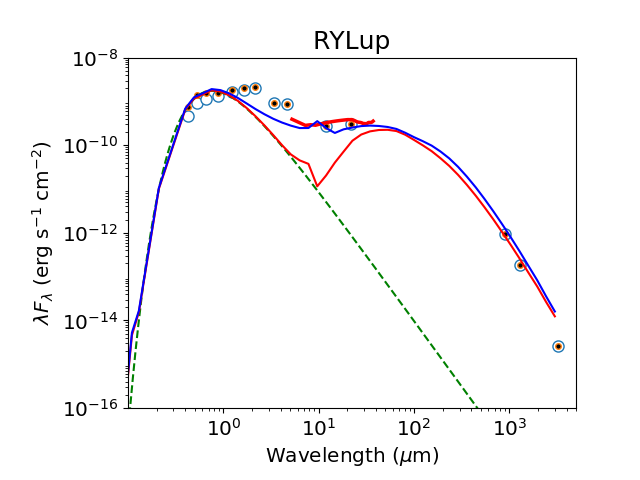}
\caption{SED model of the disk of RY Lup at 68$^{\circ}$ (red) and 38$^{\circ}$ (blue) inclination. The near infrared excess is properly reproduced at the lower inclination, whereas the ALMA images require the higher value for the outer disk. This indicates that there is a misalignment between the inner and outer disk or a warp. }
\end{center}
\label{fig:rylupmisalign}
\end{figure}

The inclination of RY~Lup is well constrained by the continuum emission ring and CO spectra at 68$^{\circ}$, but the near infrared emission in the SED cannot be reproduced if the disk is so close to edge-on. Furthermore, the $^{13}$CO emission does not show an indication of a gap in the gas, unlikely the other spatially resolved disks in our sample. One possible way to reproduce the near infrared excess is to set the inclination of the inner disk where the near infrared emission originates to 38$^{\circ}$ (or less). DALI is not capable of modeling such a misalignment between the inner and outer disk directly, but the effect is illustrated for the SED in Figure \ref{fig:rylupmisalign}. The lack of a visible $^{13}$CO gap may also be caused by this misalignment.

Another way to increase the near infrared excess is with more complex inner disk structures with a puffed up inner rim \citep{Isella2005}. Adding an artificial puffed up rim in our DALI model by increasing the scale height $h(r)$ around the sublimation radius does not reproduce the excess, although this is a rather simple approach. Although there may be other ways to reproduce the near infrared excess beyond the possibilities of DALI, we propose misalignment here as a potential solution. \citet{Manset2009} detected optical variability in this system, interpreted as occultation by a warped edge-on inner disk, suggesting that the inner disk structure may be even more complex.

The near infrared excess in the SEDs of Sz~100, J16102955 and J16070854 are also difficult to reproduce with our adopted disk model, and these disks all have high inclination angles based on their millimeter image. Based on their SED alone, they would not have been recognized as transition disks. It is striking that this issue of reproducing near infrared excess is primarily seen towards edge-on disks so we propose a misalignment to explain the near infrared excess instead. 

What could cause misalignment between inner and outer disk, also called a warp? Dynamical simulations suggest that such a misalignment can be caused by interaction with a companion in the disk \citep{Mouillet1997}. In $\beta$ Pic, the inner disk warp may have been dynamically induced by $\beta$ Pic b \citep{Dawson2011}, whose orbit is found to be aligned with the inclined warped component \citep{Chauvin2012}. This phenomenon appears to be common in TDs: for HD~142527 and HD~100453 where the inner disk warp was constrained by modeling of the shadows in the outer disk in scattered light images \citep{Marino2015, Benisty2017, Min2017}, and variations in near infrared excess and optical light curves in the face-on disk J1604-2130 \citep{vanderMarel2015-12co,Ansdell2016b} suggest a misalignment of the inner disk as well. A (planetary) companion in an inclined orbit could be responsible for the inclined inner disk \citep{OwenLai2017}. A warp can be observed directly through a deviation of Keplerian motion in the gas velocities close to the star \citep{Rosenfeld2014}, but the spatial resolution of the current CO observations is insufficient to confirm this.
Interestingly, the reason that RY~Lup was not recognized as a transition disk in previous (SED) studies, is the lack of infrared deficit in its SED photometry: a handful of other disks are known to show surprisingly large dust cavities in millimeter imaging despite a lack of infrared deficit \citep{Andrews2011}. As the inclination of the inner disk is unconstrained in most disks, it remains unclear how common this phenomenon is. 

A few other transition disks in the literature are known to have large dust cavities from millimeter imaging with a primordial SED, e.g. MWC~758, RY~Tau and WSB~60 \citep{Isella2010rytau,Andrews2011}. This raises the question whether transition disks can be recognized from SEDs alone: a face-on inner disk will produce strong near infrared excess which make the SED appear to be primordial whereas a more inclined inner disk results in a deeper infrared deficit in the SED. The amount of near infrared excess in a SED is usually fit by different amounts of dust in the inner disk ($\delta_{\rm dust}$), and it has even been suggested that there are pre-transitional and transitional disks based on the amount of near infrared excess \citep{Espaillat2011}, but perhaps this is related to orientation rather than amount. Assuming a random distribution of disk orientations, SED studies may miss a significant fraction of transition disks with large cavities. This orientation effect may be the consequence of misalignment.

\section{Conclusions}
In this work, we have quantified the dust and gas of 11 transition disks  with large cavities that were identified in the Lupus disk survey of \citet{Ansdell2016}. The analysis was performed using the physical-chemical modeling code DALI, and the modeling outcomes were compared with spatially resolved ALMA Band 7 observations of the 345 GHz continuum,  $^{13}$CO and C$^{18}$O 3--2 observations and the Spectral Energy Distribution of each source. This is the first study of transition disks using spatially resolved data within a single star forming region based on a complete disk survey. The derived gas and dust surface density profiles give information about the origin of the transition disk cavities.

\begin{enumerate}
\item The transition disks in this sample have dust cavity sizes ranging between 20 and 90 AU radius. The dust cavities are empty of millimeter dust down by at least two orders of magnitude compared with the outer disk.
\item The $^{13}$CO emission can be fit using the same surface density profile as the dust in the outer disk using a gas-to-dust ratio of 100, with various amounts of gas inside the dust cavity. The C$^{18}$O 3--2 integrated fluxes suggest that the gas-to-dust ratio is likely lower than 100 similar to previous studies.
\item For three disks, the data reveal a gas cavity smaller than the dust cavity, similar to previous studies. For the other disks the gas cavity radius cannot be constrained and is set equal to the dust cavity radius. \item The drop in density inside the gas cavity is several orders of magnitude for half of the sample, and unconstrained for the remainder of the sample. 
\item The deep drops and large cavity sizes are consistent with clearing by giant planets, but the observational exoplanet statistics of giant planets at wide orbits is three times lower than the fraction of transition disks in this  survey of $\gtrsim$11\%. Either clearing is done by (multiple) lower mass planets or subsequent migration may play an important role.
\item This study extends the sample of known transition disks with large cavities down to late spectral types (K and M).
\item Transition disks with large cavities are millimeter-bright and have large outer dust radii compared to primordial disks, consistent with higher disk masses. 
\item Transition disks with large cavities have similar accretion rates as primordial disks and are primarily found around more luminous stars.
\item The connection between large, extended ring-like disks and transition disks indicate two different evolutionary pathways for protoplanetary disks: massive, large disks go through a transition disk phase with planet clearing by giant planets, whereas smaller, low-mass disks dissipate slowly without forming large dust cavities.
\item Four disks in our sample would not have been recognized as transition disks from their photometric SED alone due to large mid-infrared excess, potentially caused by a strong silicate feature. It is suggested for at least one disk that this is caused by a misalignment between the inner and outer disk. 
\end{enumerate}

  \begin{acknowledgements}
  The authors would like to thank the referee for their helpful comments and Ruobing Dong for useful discussion.
  NM is supported by the Beatrice Watson Parrent Fellowship.
  Astrochemistry in Leiden is supported by the Netherlands
  Research School for Astronomy (NOVA), by a Royal Netherlands Academy
  of Arts and Sciences (KNAW) professor prize, and by the European
  Union A-ERC grant 291141 CHEMPLAN. This research was supported by the Munich Institute for Astro- and Particle Physics (MIAPP) of the DFG cluster of excellence "Origin and Structure of the Universe. This paper makes use of the
  following ALMA data: ADS/JAO.ALMA\#2013.1.00220.S and ADS/JAO.ALMA\#2012.1.00761.S. ALMA is a partnership of ESO (representing its member states), NSF (USA) and
  NINS (Japan), together with NRC (Canada) and NSC and ASIAA (Taiwan),
  in cooperation with the Republic of Chile. The Joint ALMA
  Observatory is operated by ESO, AUI/NRAO and NAOJ. MT has been supported by the DISCSIM project, grant agreement 341137
funded by the European Research Council under ERC-2013-ADG. CM and AM acknowledge an ESO fellowship. 
  \end{acknowledgements}

\bibliographystyle{apj}

\begin{thebibliography}{}
\expandafter\ifx\csname natexlab\endcsname\relax\def\natexlab#1{#1}\fi

\bibitem[{{Alcal{\'a}} {et~al.}(2014){Alcal{\'a}}, {Natta}, {Manara}, {Spezzi},
  {Stelzer}, {Frasca}, {Biazzo}, {Covino}, {Randich}, {Rigliaco}, {Testi},
  {Comer{\'o}n}, {Cupani}, \& {D'Elia}}]{Alcala2014}
{Alcal{\'a}}, J.~M., {Natta}, A., {Manara}, C.~F., {et~al.} 2014, \aap, 561, A2

\bibitem[{{Alcal{\'a}} {et~al.}(2017){Alcal{\'a}}, {Manara}, {Natta}, {Frasca},
  {Testi}, {Nisini}, {Stelzer}, {Williams}, {Antoniucci}, {Biazzo}, {Covino},
  {Esposito}, {Getman}, \& {Rigliaco}}]{Alcala2017}
{Alcal{\'a}}, J.~M., {Manara}, C.~F., {Natta}, A., {et~al.} 2017, \aap, 600,
  A20

\bibitem[{{Alexander} {et~al.}(2014){Alexander}, {Pascucci}, {Andrews},
  {Armitage}, \& {Cieza}}]{Alexander2014}
{Alexander}, R., {Pascucci}, I., {Andrews}, S., {Armitage}, P., \& {Cieza}, L.
  2014, Protostars and Planets VI, 475

\bibitem[{{ALMA Partnership} {et~al.}(2015){ALMA Partnership}, {Brogan},
  {P{\'e}rez}, {Hunter}, {Dent}, {Hales}, {Hills}, {Corder}, {Fomalont},
  {Vlahakis}, {Asaki}, {Barkats}, {Hirota}, {Hodge}, {Impellizzeri}, {Kneissl},
  {Liuzzo}, {Lucas}, {Marcelino}, {Matsushita}, {Nakanishi}, {Phillips},
  {Richards}, {Toledo}, {Aladro}, {Broguiere}, {Cortes}, {Cortes}, {Espada},
  {Galarza}, {Garcia-Appadoo}, {Guzman-Ramirez}, {Humphreys}, {Jung}, {Kameno},
  {Laing}, {Leon}, {Marconi}, {Mignano}, {Nikolic}, {Nyman}, {Radiszcz},
  {Remijan}, {Rod{\'o}n}, {Sawada}, {Takahashi}, {Tilanus}, {Vila Vilaro},
  {Watson}, {Wiklind}, {Akiyama}, {Chapillon}, {de Gregorio-Monsalvo}, {Di
  Francesco}, {Gueth}, {Kawamura}, {Lee}, {Nguyen Luong}, {Mangum}, {Pietu},
  {Sanhueza}, {Saigo}, {Takakuwa}, {Ubach}, {van Kempen}, {Wootten},
  {Castro-Carrizo}, {Francke}, {Gallardo}, {Garcia}, {Gonzalez}, {Hill},
  {Kaminski}, {Kurono}, {Liu}, {Lopez}, {Morales}, {Plarre}, {Schieven},
  {Testi}, {Videla}, {Villard}, {Andreani}, {Hibbard}, \&
  {Tatematsu}}]{HLTau2015}
{ALMA Partnership}, A., {Brogan}, C.~L., {P{\'e}rez}, L.~M., {et~al.} 2015,
  \apjl, 808, L3

\bibitem[{{Andrews} {et~al.}(2013){Andrews}, {Rosenfeld}, {Kraus}, \&
  {Wilner}}]{Andrews2013}
{Andrews}, S.~M., {Rosenfeld}, K.~A., {Kraus}, A.~L., \& {Wilner}, D.~J. 2013,
  \apj, 771, 129

\bibitem[{{Andrews} {et~al.}(2011){Andrews}, {Wilner}, {Espaillat}, {Hughes},
  {Dullemond}, {McClure}, {Qi}, \& {Brown}}]{Andrews2011}
{Andrews}, S.~M., {Wilner}, D.~J., {Espaillat}, C., {et~al.} 2011, \apj, 732,
  42

\bibitem[{{Ansdell} {et~al.}(2016{\natexlab{a}}){Ansdell}, {Gaidos},
  {Williams}, {Kennedy}, {Wyatt}, {LaCourse}, {Jacobs}, \&
  {Mann}}]{Ansdell2016b}
{Ansdell}, M., {Gaidos}, E., {Williams}, J.~P., {et~al.} 2016{\natexlab{a}},
  \mnras, 462, L101

\bibitem[{{Ansdell} {et~al.}(2016{\natexlab{b}}){Ansdell}, {Williams}, {van der
  Marel}, {Carpenter}, {Guidi}, {Hogerheijde}, {Mathews}, {Manara}, {Miotello},
  {Natta}, {Oliveira}, {Tazzari}, {Testi}, {van Dishoeck}, \& {van
  Terwisga}}]{Ansdell2016}
{Ansdell}, M., {Williams}, J.~P., {van der Marel}, N., {et~al.}
  2016{\natexlab{b}}, \apj, 828, 46

\bibitem[{{Armitage}(2011)}]{Armitage2011}
{Armitage}, P.~J. 2011, ARAA, 49, 195

\bibitem[{{Ataiee} {et~al.}(2013){Ataiee}, {Pinilla}, {Zsom}, {Dullemond},
  {Dominik}, \& {Ghanbari}}]{Ataiee2013}
{Ataiee}, S., {Pinilla}, P., {Zsom}, A., {et~al.} 2013, \aap, 553, L3

\bibitem[{{Barge} \& {Sommeria}(1995)}]{BargeSommeria1995}
{Barge}, P., \& {Sommeria}, J. 1995, \aap, 295, L1

\bibitem[{{Benisty} {et~al.}(2017){Benisty}, {Stolker}, {Pohl}, {de Boer},
  {Lesur}, {Dominik}, {Dullemond}, {Langlois}, {Min}, {Wagner}, {Henning},
  {Juhasz}, {Pinilla}, {Facchini}, {Apai}, {van Boekel}, {Garufi}, {Ginski},
  {M{\'e}nard}, {Pinte}, {Quanz}, {Zurlo}, {Boccaletti}, {Bonnefoy}, {Beuzit},
  {Chauvin}, {Cudel}, {Desidera}, {Feldt}, {Fontanive}, {Gratton}, {Kasper},
  {Lagrange}, {LeCoroller}, {Mouillet}, {Mesa}, {Sissa}, {Vigan}, {Antichi},
  {Buey}, {Fusco}, {Gisler}, {Llored}, {Magnard}, {Moeller-Nilsson}, {Pragt},
  {Roelfsema}, {Sauvage}, \& {Wildi}}]{Benisty2017}
{Benisty}, M., {Stolker}, T., {Pohl}, A., {et~al.} 2017, \aap, 597, A42

\bibitem[{{Biller} {et~al.}(2014){Biller}, {Males}, {Rodigas}, {Morzinski},
  {Close}, {Juh{\'a}sz}, {Follette}, {Lacour}, {Benisty}, {Sicilia-Aguilar},
  {Hinz}, {Weinberger}, {Henning}, {Pott}, {Bonnefoy}, \&
  {K{\"o}hler}}]{Biller2014}
{Biller}, B.~A., {Males}, J., {Rodigas}, T., {et~al.} 2014, \apjl, 792, L22

\bibitem[{{Boehler} {et~al.}(2017){Boehler}, {Weaver}, {Isella}, {Ricci},
  {Grady}, {Carpenter}, \& {Perez}}]{Boehler2017}
{Boehler}, Y., {Weaver}, E., {Isella}, A., {et~al.} 2017, \apj, 840, 60

\bibitem[{{Bowler}(2016)}]{Bowler2016}
{Bowler}, B.~P. 2016, \pasp, 128, 102001

\bibitem[{{Brown} {et~al.}(2009){Brown}, {Blake}, {Qi}, {Dullemond}, {Wilner},
  \& {Williams}}]{Brown2009}
{Brown}, J.~M., {Blake}, G.~A., {Qi}, C., {et~al.} 2009, \apj, 704, 496

\bibitem[{{Brown} {et~al.}(2012){Brown}, {Herczeg}, {Pontoppidan}, \& {van
  Dishoeck}}]{Brown2012a}
{Brown}, J.~M., {Herczeg}, G.~J., {Pontoppidan}, K.~M., \& {van Dishoeck},
  E.~F. 2012, \apj, 744, 116

\bibitem[{{Bruderer}(2013)}]{Bruderer2013}
{Bruderer}, S. 2013, \aap, 559, A46

\bibitem[{{Bruderer} {et~al.}(2014){Bruderer}, {van der Marel}, {van Dishoeck},
  \& {van Kempen}}]{Bruderer2014}
{Bruderer}, S., {van der Marel}, N., {van Dishoeck}, E.~F., \& {van Kempen},
  T.~A. 2014, \aap, 562, A26

\bibitem[{{Bruderer} {et~al.}(2012){Bruderer}, {van Dishoeck}, {Doty}, \&
  {Herczeg}}]{Bruderer2012}
{Bruderer}, S., {van Dishoeck}, E.~F., {Doty}, S.~D., \& {Herczeg}, G.~J. 2012,
  \aap, 541, A91

\bibitem[{{Bustamante} {et~al.}(2015){Bustamante}, {Mer{\'{\i}}n}, {Ribas},
  {Bouy}, {Prusti}, {Pilbratt}, \& {Andr{\'e}}}]{Bustamante2015}
{Bustamante}, I., {Mer{\'{\i}}n}, B., {Ribas}, {\'A}., {et~al.} 2015, \aap,
  578, A23

\bibitem[{{Calvet} {et~al.}(2002){Calvet}, {D'Alessio}, {Hartmann}, {Wilner},
  {Walsh}, \& {Sitko}}]{Calvet2002}
{Calvet}, N., {D'Alessio}, P., {Hartmann}, L., {et~al.} 2002, \apj, 568, 1008

\bibitem[{{Canovas} {et~al.}(2016){Canovas}, {Caceres}, {Schreiber}, {Hardy},
  {Cieza}, {M{\'e}nard}, \& {Hales}}]{Canovas2016}
{Canovas}, H., {Caceres}, C., {Schreiber}, M.~R., {et~al.} 2016, \mnras, 458,
  L29

\bibitem[{{Canovas} {et~al.}(2015){Canovas}, {Schreiber}, {C{\'a}ceres},
  {M{\'e}nard}, {Pinte}, {Mathews}, {Cieza}, {Casassus}, {Hales}, {Williams},
  {Rom{\'a}n}, \& {Hardy}}]{Canovas2015}
{Canovas}, H., {Schreiber}, M.~R., {C{\'a}ceres}, C., {et~al.} 2015, \apj, 805,
  21

\bibitem[{{Carmona} {et~al.}(2017){Carmona}, {Thi}, {Kamp}, {Baruteau},
  {Matter}, {van den Ancker}, {Pinte}, {K{\'o}sp{\'a}l}, {Audard}, {Liebhart},
  {Sicilia-Aguilar}, {Pinilla}, {Reg{\'a}ly}, {G{\"u}del}, {Henning}, {Cieza},
  {Baldovin-Saavedra}, {Meeus}, \& {Eiroa}}]{Carmona2017}
{Carmona}, A., {Thi}, W.~F., {Kamp}, I., {et~al.} 2017, \aap, 598, A118

\bibitem[{{Carpenter} {et~al.}(2014){Carpenter}, {Ricci}, \&
  {Isella}}]{Carpenter2014}
{Carpenter}, J.~M., {Ricci}, L., \& {Isella}, A. 2014, \apj, 787, 42

\bibitem[{{Casassus} {et~al.}(2013){Casassus}, {van der Plas}, {M}, {Dent},
  {Fomalont}, {Hagelberg}, {Hales}, {Jord{\'a}n}, {Mawet}, {M{\'e}nard},
  {Wootten}, {Wilner}, {Hughes}, {Schreiber}, {Girard}, {Ercolano}, {Canovas},
  {Rom{\'a}n}, \& {Salinas}}]{Casassus2013}
{Casassus}, S., {van der Plas}, G., {M}, S.~P., {et~al.} 2013, \nat, 493, 191

\bibitem[{{Casassus} {et~al.}(2015){Casassus}, {Wright}, {Marino}, {Maddison},
  {Wootten}, {Roman}, {P{\'e}rez}, {Pinilla}, {Wyatt}, {Moral}, {M{\'e}nard},
  {Christiaens}, {Cieza}, \& {van der Plas}}]{Casassus2015}
{Casassus}, S., {Wright}, C.~M., {Marino}, S., {et~al.} 2015, \apj, 812, 126

\bibitem[{{Chapillon} {et~al.}(2008){Chapillon}, {Guilloteau}, {Dutrey}, \&
  {Pi{\'e}tu}}]{Chapillon2008}
{Chapillon}, E., {Guilloteau}, S., {Dutrey}, A., \& {Pi{\'e}tu}, V. 2008, \aap,
  488, 565

\bibitem[{{Chauvin} {et~al.}(2012){Chauvin}, {Lagrange}, {Beust}, {Bonnefoy},
  {Boccaletti}, {Apai}, {Allard}, {Ehrenreich}, {Girard}, {Mouillet}, \&
  {Rouan}}]{Chauvin2012}
{Chauvin}, G., {Lagrange}, A.-M., {Beust}, H., {et~al.} 2012, \aap, 542, A41

\bibitem[{{Chiang} \& {Goldreich}(1997)}]{ChiangGoldreich1997}
{Chiang}, E.~I., \& {Goldreich}, P. 1997, \apj, 490, 368

\bibitem[{{Cieza} {et~al.}(2007){Cieza}, {Padgett}, {Stapelfeldt}, {Augereau},
  {Harvey}, {Evans}, {Mer{\'{\i}}n}, {Koerner}, {Sargent}, {van Dishoeck},
  {Allen}, {Blake}, {Brooke}, {Chapman}, {Huard}, {Lai}, {Mundy}, {Myers},
  {Spiesman}, \& {Wahhaj}}]{Cieza2007}
{Cieza}, L., {Padgett}, D.~L., {Stapelfeldt}, K.~R., {et~al.} 2007, \apj, 667,
  308

\bibitem[{{Cieza} {et~al.}(2012){Cieza}, {Schreiber}, {Romero}, {Williams},
  {Rebassa-Mansergas}, \& {Mer{\'{\i}}n}}]{Cieza2012}
{Cieza}, L.~A., {Schreiber}, M.~R., {Romero}, G.~A., {et~al.} 2012, \apj, 750,
  157

\bibitem[{{Cieza} {et~al.}(2010){Cieza}, {Schreiber}, {Romero}, {Mora},
  {Merin}, {Swift}, {Orellana}, {Williams}, {Harvey}, \& {Evans}}]{Cieza2010}
---. 2010, \apj, 712, 925

\bibitem[{{Currie} {et~al.}(2015){Currie}, {Cloutier}, {Brittain}, {Grady},
  {Burrows}, {Muto}, {Kenyon}, \& {Kuchner}}]{Currie2015}
{Currie}, T., {Cloutier}, R., {Brittain}, S., {et~al.} 2015, \apjl, 814, L27

\bibitem[{{Currie} \& {Kenyon}(2009)}]{Currie2009}
{Currie}, T., \& {Kenyon}, S.~J. 2009, \aj, 138, 703

\bibitem[{{Dawson} {et~al.}(2011){Dawson}, {Murray-Clay}, \&
  {Fabrycky}}]{Dawson2011}
{Dawson}, R.~I., {Murray-Clay}, R.~A., \& {Fabrycky}, D.~C. 2011, \apjl, 743,
  L17

\bibitem[{{Dong} {et~al.}(2017){Dong}, {van der Marel}, {Hashimoto}, {Chiang},
  {Akiyama}, {Liu}, {Muto}, {Knapp}, {Tsukagoshi}, {Brown}, {Bruderer},
  {Koyamatsu}, {Kudo}, {Ohashi}, {Rich}, {Satoshi}, {Takami}, {Wisniewski},
  {Yang}, {Zhu}, \& {Tamura}}]{Dong2017}
{Dong}, R., {van der Marel}, N., {Hashimoto}, J., {et~al.} 2017, \apj, 836, 201

\bibitem[{{Dunkin} {et~al.}(1997){Dunkin}, {Barlow}, \& {Ryan}}]{Dunkin1997}
{Dunkin}, S.~K., {Barlow}, M.~J., \& {Ryan}, S.~G. 1997, \mnras, 290, 165

\bibitem[{{Dutrey} {et~al.}(2008){Dutrey}, {Guilloteau}, {Pi{\'e}tu},
  {Chapillon}, {Gueth}, {Henning}, {Launhardt}, {Pavlyuchenkov}, {Schreyer}, \&
  {Semenov}}]{Dutrey2008}
{Dutrey}, A., {Guilloteau}, S., {Pi{\'e}tu}, V., {et~al.} 2008, \aap, 490, L15

\bibitem[{{Ercolano} \& {Pascucci}(2017)}]{Ercolano2017}
{Ercolano}, B., \& {Pascucci}, I. 2017, Royal Society Open Science, 4, 170114

\bibitem[{{Ercolano} {et~al.}(2017){Ercolano}, {Weber}, \&
  {Owen}}]{Ercolano2017xray}
{Ercolano}, B., {Weber}, M.~L., \& {Owen}, J.~E. 2017, ArXiv e-prints,
  arXiv:1710.03816

\bibitem[{{Espaillat} {et~al.}(2011){Espaillat}, {Furlan}, {D'Alessio},
  {Sargent}, {Nagel}, {Calvet}, {Watson}, \& {Muzerolle}}]{Espaillat2011}
{Espaillat}, C., {Furlan}, E., {D'Alessio}, P., {et~al.} 2011, \apj, 728, 49

\bibitem[{{Espaillat} {et~al.}(2014){Espaillat}, {Muzerolle}, {Najita},
  {Andrews}, {Zhu}, {Calvet}, {Kraus}, {Hashimoto}, {Kraus}, \&
  {D'Alessio}}]{Espaillat2014}
{Espaillat}, C., {Muzerolle}, J., {Najita}, J., {et~al.} 2014, Protostars and
  Planets VI, 497

\bibitem[{{Evans} {et~al.}(2009){Evans}, {Dunham}, {J{\o}rgensen}, {Enoch},
  {Mer{\'{\i}}n}, {van Dishoeck}, {Alcal{\'a}}, {Myers}, {Stapelfeldt},
  {Huard}, {Allen}, {Harvey}, {van Kempen}, {Blake}, {Koerner}, {Mundy},
  {Padgett}, \& {Sargent}}]{Evans2009}
{Evans}, II, N.~J., {Dunham}, M.~M., {J{\o}rgensen}, J.~K., {et~al.} 2009,
  \apjs, 181, 321

\bibitem[{{Facchini} {et~al.}(2017){Facchini}, {Birnstiel}, {Bruderer}, \& {van
  Dishoeck}}]{Facchini2017gaps}
{Facchini}, S., {Birnstiel}, T., {Bruderer}, S., \& {van Dishoeck}, E.~F. 2017,
  subm. to A\&A

\bibitem[{{Fedele} {et~al.}(2017){Fedele}, {Carney}, {Hogerheijde}, {Walsh},
  {Miotello}, {Klaassen}, {Bruderer}, {Henning}, \& {van
  Dishoeck}}]{Fedele2017}
{Fedele}, D., {Carney}, M., {Hogerheijde}, M.~R., {et~al.} 2017, \aap, 600, A72

\bibitem[{{Fung} {et~al.}(2014){Fung}, {Shi}, \& {Chiang}}]{Fung2014}
{Fung}, J., {Shi}, J.-M., \& {Chiang}, E. 2014, \apj, 782, 88

\bibitem[{{Garcia Lopez} {et~al.}(2006){Garcia Lopez}, {Natta}, {Testi}, \&
  {Habart}}]{GarciaLopez2006}
{Garcia Lopez}, R., {Natta}, A., {Testi}, L., \& {Habart}, E. 2006, \aap, 459,
  837

\bibitem[{{Garufi} {et~al.}(2017){Garufi}, {Meeus}, {Benisty}, {Quanz},
  {Banzatti}, {Kama}, {Canovas}, {Eiroa}, {Schmid}, {Stolker}, {Pohl},
  {Rigliaco}, {M{\'e}nard}, {Meyer}, {van Boekel}, \& {Dominik}}]{Garufi2017}
{Garufi}, A., {Meeus}, G., {Benisty}, M., {et~al.} 2017, \aap, 603, A21

\bibitem[{{Hartmann} {et~al.}(1998){Hartmann}, {Calvet}, {Gullbring}, \&
  {D'Alessio}}]{Hartmann1998}
{Hartmann}, L., {Calvet}, N., {Gullbring}, E., \& {D'Alessio}, P. 1998, \apj,
  495, 385

\bibitem[{{Hern{\'a}ndez} {et~al.}(2004){Hern{\'a}ndez}, {Calvet},
  {Brice{\~n}o}, {Hartmann}, \& {Berlind}}]{Hernandez2004}
{Hern{\'a}ndez}, J., {Calvet}, N., {Brice{\~n}o}, C., {Hartmann}, L., \&
  {Berlind}, P. 2004, \aj, 127, 1682

\bibitem[{{Hu{\'e}lamo} {et~al.}(2015){Hu{\'e}lamo}, {de Gregorio-Monsalvo},
  {Macias}, {Pinte}, {Ireland}, {Tuthill}, \& {Lacour}}]{Huelamo2015}
{Hu{\'e}lamo}, N., {de Gregorio-Monsalvo}, I., {Macias}, E., {et~al.} 2015,
  \aap, 575, L5

\bibitem[{{Irvine} \& {Houk}(1977)}]{Irvine1977}
{Irvine}, N.~J., \& {Houk}, N. 1977, \pasp, 89, 347

\bibitem[{{Isella} {et~al.}(2010){Isella}, {Carpenter}, \&
  {Sargent}}]{Isella2010rytau}
{Isella}, A., {Carpenter}, J.~M., \& {Sargent}, A.~I. 2010, \apj, 714, 1746

\bibitem[{{Isella} \& {Natta}(2005)}]{Isella2005}
{Isella}, A., \& {Natta}, A. 2005, \aap, 438, 899

\bibitem[{{Isella} {et~al.}(2012){Isella}, {P{\'e}rez}, \&
  {Carpenter}}]{Isella2012}
{Isella}, A., {P{\'e}rez}, L.~M., \& {Carpenter}, J.~M. 2012, \apj, 747, 136

\bibitem[{{Isella} {et~al.}(2016){Isella}, {Guidi}, {Testi}, {Liu}, {Li}, {Li},
  {Weaver}, {Boehler}, {Carperter}, {De Gregorio-Monsalvo}, {Manara}, {Natta},
  {P{\'e}rez}, {Ricci}, {Sargent}, {Tazzari}, \& {Turner}}]{Isella2016}
{Isella}, A., {Guidi}, G., {Testi}, L., {et~al.} 2016, Physical Review Letters,
  117, 251101

\bibitem[{{Kessler-Silacci} {et~al.}(2006){Kessler-Silacci}, {Augereau},
  {Dullemond}, {Geers}, {Lahuis}, {Evans}, {van Dishoeck}, {Blake}, {Boogert},
  {Brown}, {J{\o}rgensen}, {Knez}, \& {Pontoppidan}}]{Kessler2006}
{Kessler-Silacci}, J., {Augereau}, J.-C., {Dullemond}, C.~P., {et~al.} 2006,
  \apj, 639, 275

\bibitem[{{Klahr} \& {Henning}(1997)}]{KlahrHenning1997}
{Klahr}, H.~H., \& {Henning}, T. 1997, Icarus, 128, 213

\bibitem[{{Kley} \& {Nelson}(2012)}]{KleyNelson2012}
{Kley}, W., \& {Nelson}, R.~P. 2012, \araa, 50, 211

\bibitem[{{Kraus} \& {Ireland}(2012)}]{KrausIreland2012}
{Kraus}, A.~L., \& {Ireland}, M.~J. 2012, \apj, 745, 5

\bibitem[{{Lawson} {et~al.}(2004){Lawson}, {Lyo}, \& {Muzerolle}}]{Lawson2004}
{Lawson}, W.~A., {Lyo}, A.-R., \& {Muzerolle}, J. 2004, \mnras, 351, L39

\bibitem[{{Lin} \& {Papaloizou}(1979)}]{LinPapaloizou1979}
{Lin}, D.~N.~C., \& {Papaloizou}, J. 1979, \mnras, 188, 191

\bibitem[{{Luhman}(2000)}]{Luhman2000}
{Luhman}, K.~L. 2000, \apj, 544, 1044

\bibitem[{{Luhman}(2012)}]{Luhman2012}
---. 2012, \araa, 50, 65

\bibitem[{{Lynden-Bell} \& {Pringle}(1974)}]{LyndenPringle1974}
{Lynden-Bell}, D., \& {Pringle}, J.~E. 1974, \mnras, 168, 603

\bibitem[{{Maaskant} {et~al.}(2013){Maaskant}, {Honda}, {Waters}, {Tielens},
  {Dominik}, {Min}, {Verhoeff}, {Meeus}, \& {van den Ancker}}]{Maaskant2013}
{Maaskant}, K.~M., {Honda}, M., {Waters}, L.~B.~F.~M., {et~al.} 2013, \aap,
  555, A64

\bibitem[{{Maire} {et~al.}(2017){Maire}, {Stolker}, {Messina}, {M{\"u}ller},
  {Biller}, {Currie}, {Dominik}, {Grady}, {Boccaletti}, {Bonnefoy}, {Chauvin},
  {Galicher}, {Millward}, {Pohl}, {Brandner}, {Henning}, {Lagrange},
  {Langlois}, {Meyer}, {Quanz}, {Vigan}, {Zurlo}, {van Boekel}, {Buenzli},
  {Buey}, {Desidera}, {Feldt}, {Fusco}, {Ginski}, {Giro}, {Gratton}, {Hubin},
  {Lannier}, {Le Mignant}, {Mesa}, {Peretti}, {Perrot}, {Ramos}, {Salter},
  {Samland}, {Sissa}, {Stadler}, {Thalmann}, {Udry}, \& {Weber}}]{Maire2017}
{Maire}, A.-L., {Stolker}, T., {Messina}, S., {et~al.} 2017, \aap, 601, A134

\bibitem[{{Manara} {et~al.}(2014){Manara}, {Testi}, {Natta}, {Rosotti},
  {Benisty}, {Ercolano}, \& {Ricci}}]{Manara2014}
{Manara}, C.~F., {Testi}, L., {Natta}, A., {et~al.} 2014, \aap, 568, A18

\bibitem[{{Manara} {et~al.}(2016){Manara}, {Rosotti}, {Testi}, {Natta},
  {Alcal{\'a}}, {Williams}, {Ansdell}, {Miotello}, {van der Marel}, {Tazzari},
  {Carpenter}, {Guidi}, {Mathews}, {Oliveira}, {Prusti}, \& {van
  Dishoeck}}]{Manara2016}
{Manara}, C.~F., {Rosotti}, G., {Testi}, L., {et~al.} 2016, \aap, 591, L3

\bibitem[{{Manset} {et~al.}(2009){Manset}, {Bastien}, {M{\'e}nard}, {Bertout},
  {Le van Suu}, \& {Boivin}}]{Manset2009}
{Manset}, N., {Bastien}, P., {M{\'e}nard}, F., {et~al.} 2009, \aap, 499, 137

\bibitem[{{Marino} {et~al.}(2015){Marino}, {Perez}, \& {Casassus}}]{Marino2015}
{Marino}, S., {Perez}, S., \& {Casassus}, S. 2015, \apjl, 798, L44

\bibitem[{{Mathews} {et~al.}(2012){Mathews}, {Williams}, \&
  {M{\'e}nard}}]{Mathews2012}
{Mathews}, G.~S., {Williams}, J.~P., \& {M{\'e}nard}, F. 2012, \apj, 753, 59

\bibitem[{{Mer{\'{\i}}n} {et~al.}(2004){Mer{\'{\i}}n}, {Montesinos}, {Eiroa},
  {Solano}, {Mora}, {D'Alessio}, {Calvet}, {Oudmaijer}, {de Winter}, {Davies},
  {Harris}, {Collier Cameron}, {Deeg}, {Ferlet}, {Garz{\'o}n}, {Grady},
  {Horne}, {Miranda}, {Palacios}, {Penny}, {Quirrenbach}, {Rauer}, {Schneider},
  \& {Wesselius}}]{Merin2004}
{Mer{\'{\i}}n}, B., {Montesinos}, B., {Eiroa}, C., {et~al.} 2004, \aap, 419,
  301

\bibitem[{{Mer{\'{\i}}n} {et~al.}(2008){Mer{\'{\i}}n}, {J{\o}rgensen},
  {Spezzi}, {Alcal{\'a}}, {Evans}, {Harvey}, {Prusti}, {Chapman}, {Huard}, {van
  Dishoeck}, \& {Comer{\'o}n}}]{Merin2008}
{Mer{\'{\i}}n}, B., {J{\o}rgensen}, J., {Spezzi}, L., {et~al.} 2008, \apjs,
  177, 551

\bibitem[{{Mer{\'{\i}}n} {et~al.}(2010){Mer{\'{\i}}n}, {Brown}, {Oliveira},
  {Herczeg}, {van Dishoeck}, {Bottinelli}, {Evans}, {Cieza}, {Spezzi},
  {Alcal{\'a}}, {Harvey}, {Blake}, {Bayo}, {Geers}, {Lahuis}, {Prusti},
  {Augereau}, {Olofsson}, {Walter}, \& {Chiu}}]{Merin2010}
{Mer{\'{\i}}n}, B., {Brown}, J.~M., {Oliveira}, I., {et~al.} 2010, \apj, 718,
  1200

\bibitem[{{Min} {et~al.}(2017){Min}, {Stolker}, {Dominik}, \&
  {Benisty}}]{Min2017}
{Min}, M., {Stolker}, T., {Dominik}, C., \& {Benisty}, M. 2017, ArXiv e-prints,
  arXiv:1704.01844

\bibitem[{{Miotello} {et~al.}(2014){Miotello}, {Bruderer}, \& {van
  Dishoeck}}]{Miotello2014}
{Miotello}, A., {Bruderer}, S., \& {van Dishoeck}, E.~F. 2014, \aap, 572, A96

\bibitem[{{Miotello} {et~al.}(2017){Miotello}, {van Dishoeck}, {Williams},
  {Ansdell}, {Guidi}, {Hogerheijde}, {Manara}, {Tazzari}, {Testi}, {van der
  Marel}, \& {van Terwisga}}]{Miotello2017}
{Miotello}, A., {van Dishoeck}, E.~F., {Williams}, J.~P., {et~al.} 2017, \aap,
  599, A113

\bibitem[{{Mouillet} {et~al.}(1997){Mouillet}, {Larwood}, {Papaloizou}, \&
  {Lagrange}}]{Mouillet1997}
{Mouillet}, D., {Larwood}, J.~D., {Papaloizou}, J.~C.~B., \& {Lagrange}, A.~M.
  1997, \mnras, 292, 896

\bibitem[{{Najita} {et~al.}(2015){Najita}, {Andrews}, \&
  {Muzerolle}}]{Najita2015}
{Najita}, J.~R., {Andrews}, S.~M., \& {Muzerolle}, J. 2015, \mnras, 450, 3559

\bibitem[{{Najita} {et~al.}(2007){Najita}, {Strom}, \&
  {Muzerolle}}]{Najita2007}
{Najita}, J.~R., {Strom}, S.~E., \& {Muzerolle}, J. 2007, \mnras, 378, 369

\bibitem[{{Owen} \& {Clarke}(2012)}]{OwenClarke2012}
{Owen}, J.~E., \& {Clarke}, C.~J. 2012, \mnras, 426, L96

\bibitem[{{Owen} {et~al.}(2011){Owen}, {Ercolano}, \& {Clarke}}]{Owen2011}
{Owen}, J.~E., {Ercolano}, B., \& {Clarke}, C.~J. 2011, \mnras, 412, 13

\bibitem[{{Owen} {et~al.}(2017){Owen}, {Ercolano}, \&
  {Clarke}}]{Owen2011erratum}
---. 2017, \mnras, 472, 2955

\bibitem[{{Owen} \& {Lai}(2017)}]{OwenLai2017}
{Owen}, J.~E., \& {Lai}, D. 2017, \mnras, 469, 2834

\bibitem[{{Pascucci} {et~al.}(2016){Pascucci}, {Testi}, {Herczeg}, {Long},
  {Manara}, {Hendler}, {Mulders}, {Krijt}, {Ciesla}, {Henning}, {Mohanty},
  {Drabek-Maunder}, {Apai}, {Sz{\H u}cs}, {Sacco}, \&
  {Olofsson}}]{Pascucci2016}
{Pascucci}, I., {Testi}, L., {Herczeg}, G.~J., {et~al.} 2016, \apj, 831, 125

\bibitem[{{P{\'e}rez} {et~al.}(2014){P{\'e}rez}, {Isella}, {Carpenter}, \&
  {Chandler}}]{Perez2014}
{P{\'e}rez}, L.~M., {Isella}, A., {Carpenter}, J.~M., \& {Chandler}, C.~J.
  2014, \apjl, 783, L13

\bibitem[{{P{\'e}rez} {et~al.}(2012){P{\'e}rez}, {Carpenter}, {Chandler},
  {Isella}, {Andrews}, {Ricci}, {Calvet}, {Corder}, {Deller}, {Dullemond},
  {Greaves}, {Harris}, {Henning}, {Kwon}, {Lazio}, {Linz}, {Mundy}, {Sargent},
  {Storm}, {Testi}, \& {Wilner}}]{Perez2012}
{P{\'e}rez}, L.~M., {Carpenter}, J.~M., {Chandler}, C.~J., {et~al.} 2012,
  \apjl, 760, L17

\bibitem[{{Perez} {et~al.}(2015){Perez}, {Casassus}, {M{\'e}nard}, {Roman},
  {van der Plas}, {Cieza}, {Pinte}, {Christiaens}, \& {Hales}}]{SPerez2015}
{Perez}, S., {Casassus}, S., {M{\'e}nard}, F., {et~al.} 2015, \apj, 798, 85

\bibitem[{{Pi{\'e}tu} {et~al.}(2005){Pi{\'e}tu}, {Guilloteau}, \&
  {Dutrey}}]{Pietu2005}
{Pi{\'e}tu}, V., {Guilloteau}, S., \& {Dutrey}, A. 2005, \aap, 443, 945

\bibitem[{{Pinilla} {et~al.}(2012){Pinilla}, {Benisty}, \&
  {Birnstiel}}]{Pinilla2012b}
{Pinilla}, P., {Benisty}, M., \& {Birnstiel}, T. 2012, \aap, 545, A81

\bibitem[{{Pinilla} {et~al.}(2016){Pinilla}, {Flock}, {Ovelar}, \&
  {Birnstiel}}]{Pinilla2016dz}
{Pinilla}, P., {Flock}, M., {Ovelar}, M.~d.~J., \& {Birnstiel}, T. 2016, \aap,
  596, A81

\bibitem[{{Pinilla} {et~al.}(2015){Pinilla}, {van der Marel}, {P{\'e}rez}, {van
  Dishoeck}, {Andrews}, {Birnstiel}, {Herczeg}, {Pontoppidan}, \& {van
  Kempen}}]{Pinilla2015beta}
{Pinilla}, P., {van der Marel}, N., {P{\'e}rez}, L.~M., {et~al.} 2015, \aap,
  584, A16

\bibitem[{{Pinilla} {et~al.}(2017){Pinilla}, {P{\'e}rez}, {Andrews}, {van der
  Marel}, {van Dishoeck}, {Ataiee}, {Benisty}, {Birnstiel}, {Juh{\'a}sz},
  {Natta}, {Ricci}, \& {Testi}}]{Pinilla2017sr24s}
{Pinilla}, P., {P{\'e}rez}, L.~M., {Andrews}, S., {et~al.} 2017, \apj, 839, 99

\bibitem[{{Pohl} {et~al.}(2017){Pohl}, {Sissa}, {Langlois}, {M{\"u}ller},
  {Ginski}, {van Holstein}, {Vigan}, {Mesa}, {Maire}, {Henning}, {Gratton},
  {Olofsson}, {van Boekel}, {Benisty}, {Biller}, {Boccaletti}, {Chauvin},
  {Daemgen}, {de Boer}, {Desidera}, {Dominik}, {Garufi}, {Janson}, {Kral},
  {M{\'e}nard}, {Pinte}, {Stolker}, {Szul{\'a}gyi}, {Zurlo}, {Bonnefoy},
  {Cheetham}, {Cudel}, {Feldt}, {Kasper}, {Lagrange}, {Perrot}, \&
  {Wildi}}]{Pohl2017}
{Pohl}, A., {Sissa}, E., {Langlois}, M., {et~al.} 2017, \aap, 605, A34

\bibitem[{{Pontoppidan} {et~al.}(2008){Pontoppidan}, {Blake}, {van Dishoeck},
  {Smette}, {Ireland}, \& {Brown}}]{Pontoppidan2008}
{Pontoppidan}, K.~M., {Blake}, G.~A., {van Dishoeck}, E.~F., {et~al.} 2008,
  \apj, 684, 1323

\bibitem[{{Quanz}(2015)}]{Quanz2015}
{Quanz}, S.~P. 2015, \apss, 357, 148

\bibitem[{{Quanz} {et~al.}(2013){Quanz}, {Amara}, {Meyer}, {Kenworthy},
  {Kasper}, \& {Girard}}]{Quanz2013}
{Quanz}, S.~P., {Amara}, A., {Meyer}, M.~R., {et~al.} 2013, \apjl, 766, L1

\bibitem[{{Ragusa} {et~al.}(2017){Ragusa}, {Dipierro}, {Lodato}, {Laibe}, \&
  {Price}}]{Ragusa2017}
{Ragusa}, E., {Dipierro}, G., {Lodato}, G., {Laibe}, G., \& {Price}, D.~J.
  2017, \mnras, 464, 1449

\bibitem[{{Reg{\'a}ly} {et~al.}(2012){Reg{\'a}ly}, {Juh{\'a}sz}, {S{\'a}ndor},
  \& {Dullemond}}]{Regaly2012}
{Reg{\'a}ly}, Z., {Juh{\'a}sz}, A., {S{\'a}ndor}, Z., \& {Dullemond}, C.~P.
  2012, \mnras, 419, 1701

\bibitem[{{Reggiani} {et~al.}(2014){Reggiani}, {Quanz}, {Meyer}, {Pueyo},
  {Absil}, {Amara}, {Anglada}, {Avenhaus}, {Girard}, {Carrasco Gonzalez},
  {Graham}, {Mawet}, {Meru}, {Milli}, {Osorio}, {Wolff}, \&
  {Torrelles}}]{Reggiani2014}
{Reggiani}, M., {Quanz}, S.~P., {Meyer}, M.~R., {et~al.} 2014, \apjl, 792, L23

\bibitem[{{Riaud} {et~al.}(2006){Riaud}, {Mawet}, {Absil}, {Boccaletti},
  {Baudoz}, {Herwats}, \& {Surdej}}]{Riaud2006}
{Riaud}, P., {Mawet}, D., {Absil}, O., {et~al.} 2006, \aap, 458, 317

\bibitem[{{Rosenfeld} {et~al.}(2014){Rosenfeld}, {Chiang}, \&
  {Andrews}}]{Rosenfeld2014}
{Rosenfeld}, K.~A., {Chiang}, E., \& {Andrews}, S.~M. 2014, \apj, 782, 62

\bibitem[{{Rosotti} {et~al.}(2013){Rosotti}, {Ercolano}, {Owen}, \&
  {Armitage}}]{Rosotti2013}
{Rosotti}, G.~P., {Ercolano}, B., {Owen}, J.~E., \& {Armitage}, P.~J. 2013,
  \mnras, 430, 1392

\bibitem[{{Sallum} {et~al.}(2015){Sallum}, {Follette}, {Eisner}, {Close},
  {Hinz}, {Kratter}, {Males}, {Skemer}, {Macintosh}, {Tuthill}, {Bailey},
  {Defr{\`e}re}, {Morzinski}, {Rodigas}, {Spalding}, {Vaz}, \&
  {Weinberger}}]{Sallum2015}
{Sallum}, S., {Follette}, K.~B., {Eisner}, J.~A., {et~al.} 2015, \nat, 527, 342

\bibitem[{{Schisano} {et~al.}(2009){Schisano}, {Covino}, {Alcal{\'a}},
  {Esposito}, {Gandolfi}, \& {Guenther}}]{Schisano2009}
{Schisano}, E., {Covino}, E., {Alcal{\'a}}, J.~M., {et~al.} 2009, \aap, 501,
  1013

\bibitem[{{Siess} {et~al.}(2000){Siess}, {Dufour}, \& {Forestini}}]{Siess2000}
{Siess}, L., {Dufour}, E., \& {Forestini}, M. 2000, \aap, 358, 593

\bibitem[{{Spezzi} {et~al.}(2008){Spezzi}, {Alcal{\'a}}, {Covino}, {Frasca},
  {Gandolfi}, {Oliveira}, {Chapman}, {Evans}, {Huard}, {J{\o}rgensen},
  {Mer{\'{\i}}n}, \& {Stapelfeldt}}]{Spezzi2008}
{Spezzi}, L., {Alcal{\'a}}, J.~M., {Covino}, E., {et~al.} 2008, \apj, 680, 1295

\bibitem[{{Strom} {et~al.}(1989){Strom}, {Strom}, {Edwards}, {Cabrit}, \&
  {Skrutskie}}]{Strom1989}
{Strom}, K.~M., {Strom}, S.~E., {Edwards}, S., {Cabrit}, S., \& {Skrutskie},
  M.~F. 1989, \aj, 97, 1451

\bibitem[{{Tazzari} {et~al.}(2016){Tazzari}, {Testi}, {Ercolano}, {Natta},
  {Isella}, {Chandler}, {P{\'e}rez}, {Andrews}, {Wilner}, {Ricci}, {Henning},
  {Linz}, {Kwon}, {Corder}, {Dullemond}, {Carpenter}, {Sargent}, {Mundy},
  {Storm}, {Calvet}, {Greaves}, {Lazio}, \& {Deller}}]{Tazzari2016}
{Tazzari}, M., {Testi}, L., {Ercolano}, B., {et~al.} 2016, \aap, 588, A53

\bibitem[{{Tazzari} {et~al.}(2017){Tazzari}, {Testi}, {Natta}, {Ansdell},
  {Carpenter}, {Guidi}, {Hogerheijde}, {Manara}, {Miotello}, {van der Marel},
  {van Dishoeck}, \& {Williams}}]{Tazzari2017}
{Tazzari}, M., {Testi}, L., {Natta}, A., {et~al.} 2017, ArXiv e-prints,
  arXiv:1707.01499

\bibitem[{{Thalmann} {et~al.}(2016){Thalmann}, {Janson}, {Garufi},
  {Boccaletti}, {Quanz}, {Sissa}, {Gratton}, {Salter}, {Benisty}, {Bonnefoy},
  {Chauvin}, {Daemgen}, {Desidera}, {Dominik}, {Engler}, {Feldt}, {Henning},
  {Lagrange}, {Langlois}, {Lannier}, {Le Coroller}, {Ligi}, {M{\'e}nard},
  {Mesa}, {Meyer}, {Mulders}, {Olofsson}, {Pinte}, {Schmid}, {Vigan}, \&
  {Zurlo}}]{Thalmann2016}
{Thalmann}, C., {Janson}, M., {Garufi}, A., {et~al.} 2016, \apjl, 828, L17

\bibitem[{{Turner} {et~al.}(2014){Turner}, {Fromang}, {Gammie}, {Klahr},
  {Lesur}, {Wardle}, \& {Bai}}]{Turner2014}
{Turner}, N.~J., {Fromang}, S., {Gammie}, C., {et~al.} 2014, Protostars and
  Planets VI, 411

\bibitem[{{van der Marel} {et~al.}(2015{\natexlab{a}}){van der Marel},
  {Pinilla}, {Tobin}, {van Kempen}, {Andrews}, {Ricci}, \&
  {Birnstiel}}]{vanderMarel2015vla}
{van der Marel}, N., {Pinilla}, P., {Tobin}, J., {et~al.} 2015{\natexlab{a}},
  \apjl, 810, L7

\bibitem[{{van der Marel} {et~al.}(2016{\natexlab{a}}){van der Marel}, {van
  Dishoeck}, {Bruderer}, {Andrews}, {Pontoppidan}, {Herczeg}, {van Kempen}, \&
  {Miotello}}]{vanderMarel2016-isot}
{van der Marel}, N., {van Dishoeck}, E.~F., {Bruderer}, S., {et~al.}
  2016{\natexlab{a}}, \aap, 585, A58

\bibitem[{{van der Marel} {et~al.}(2015{\natexlab{b}}){van der Marel}, {van
  Dishoeck}, {Bruderer}, {P{\'e}rez}, \& {Isella}}]{vanderMarel2015-12co}
{van der Marel}, N., {van Dishoeck}, E.~F., {Bruderer}, S., {P{\'e}rez}, L., \&
  {Isella}, A. 2015{\natexlab{b}}, \aap, 579, A106

\bibitem[{{van der Marel} {et~al.}(2016{\natexlab{b}}){van der Marel},
  {Verhaar}, {van Terwisga}, {Mer{\'{\i}}n}, {Herczeg}, {Ligterink}, \& {van
  Dishoeck}}]{vanderMarel2016-spitzer}
{van der Marel}, N., {Verhaar}, B.~W., {van Terwisga}, S., {et~al.}
  2016{\natexlab{b}}, \aap, 592, A126

\bibitem[{{van der Marel} {et~al.}(2013){van der Marel}, {van Dishoeck},
  {Bruderer}, {Birnstiel}, {Pinilla}, {Dullemond}, {van Kempen}, {Schmalzl},
  {Brown}, {Herczeg}, {Mathews}, \& {Geers}}]{vanderMarel2013}
{van der Marel}, N., {van Dishoeck}, E.~F., {Bruderer}, S., {et~al.} 2013,
  Science, 340, 1199

\bibitem[{{van der Plas} {et~al.}(2017{\natexlab{a}}){van der Plas},
  {M{\'e}nard}, {Canovas}, {Avenhaus}, {Casassus}, {Pinte}, {Caceres}, \&
  {Cieza}}]{vanderPlas2017}
{van der Plas}, G., {M{\'e}nard}, F., {Canovas}, H., {et~al.}
  2017{\natexlab{a}}, \aap, 607, A55

\bibitem[{{van der Plas} {et~al.}(2017{\natexlab{b}}){van der Plas}, {Wright},
  {M{\'e}nard}, {Casassus}, {Canovas}, {Pinte}, {Maddison}, {Maaskant},
  {Avenhaus}, {Cieza}, {Perez}, \& {Ubach}}]{vanderPlas2016}
{van der Plas}, G., {Wright}, C.~M., {M{\'e}nard}, F., {et~al.}
  2017{\natexlab{b}}, \aap, 597, A32

\bibitem[{{Vieira} {et~al.}(2003){Vieira}, {Corradi}, {Alencar}, {Mendes},
  {Torres}, {Quast}, {Guimar{\~a}es}, \& {da Silva}}]{Vieira2003}
{Vieira}, S.~L.~A., {Corradi}, W.~J.~B., {Alencar}, S.~H.~P., {et~al.} 2003,
  \aj, 126, 2971

\bibitem[{{Wahhaj} {et~al.}(2010){Wahhaj}, {Cieza}, {Koerner}, {Stapelfeldt},
  {Padgett}, {Case}, {Keller}, {Mer{\'{\i}}n}, {Evans}, {Harvey}, {Sargent},
  {van Dishoeck}, {Allen}, {Blake}, {Brooke}, {Chapman}, {Mundy}, \&
  {Myers}}]{Wahhaj2010}
{Wahhaj}, Z., {Cieza}, L., {Koerner}, D.~W., {et~al.} 2010, \apj, 724, 835

\bibitem[{{Walsh} {et~al.}(2014){Walsh}, {Juh{\'a}sz}, {Pinilla}, {Harsono},
  {Mathews}, {Dent}, {Hogerheijde}, {Birnstiel}, {Meeus}, {Nomura}, {Aikawa},
  {Millar}, \& {Sandell}}]{Walsh2014}
{Walsh}, C., {Juh{\'a}sz}, A., {Pinilla}, P., {et~al.} 2014, \apjl, 791, L6

\bibitem[{{Wang} \& {Goodman}(2017)}]{WangGoodman2017}
{Wang}, L., \& {Goodman}, J.~J. 2017, \apj, 835, 59

\bibitem[{{Weidenschilling}(1977)}]{Weidenschilling1977}
{Weidenschilling}, S.~J. 1977, \mnras, 180, 57

\bibitem[{{Williams} \& {Cieza}(2011)}]{WilliamsCieza2011}
{Williams}, J.~P., \& {Cieza}, L.~A. 2011, \araa, 49, 67

\bibitem[{{Wu} \& {Murray}(2003)}]{WuMurray2003}
{Wu}, Y., \& {Murray}, N. 2003, \apj, 589, 605

\bibitem[{{Wyatt}(2008)}]{Wyatt2008}
{Wyatt}, M.~C. 2008, \araa, 46, 339

\bibitem[{{Zhang} {et~al.}(2014){Zhang}, {Isella}, {Carpenter}, \&
  {Blake}}]{Zhang2014}
{Zhang}, K., {Isella}, A., {Carpenter}, J.~M., \& {Blake}, G.~A. 2014, \apj,
  791, 42

\bibitem[{{Zhu} {et~al.}(2011){Zhu}, {Nelson}, {Hartmann}, {Espaillat}, \&
  {Calvet}}]{Zhu2011}
{Zhu}, Z., {Nelson}, R.~P., {Hartmann}, L., {Espaillat}, C., \& {Calvet}, N.
  2011, \apj, 729, 47

\end{thebibliography}

\appendix
\section{Additional models}

\begin{figure*}[!ht]
\includegraphics[width=\textwidth, trim=150 100 150 150]{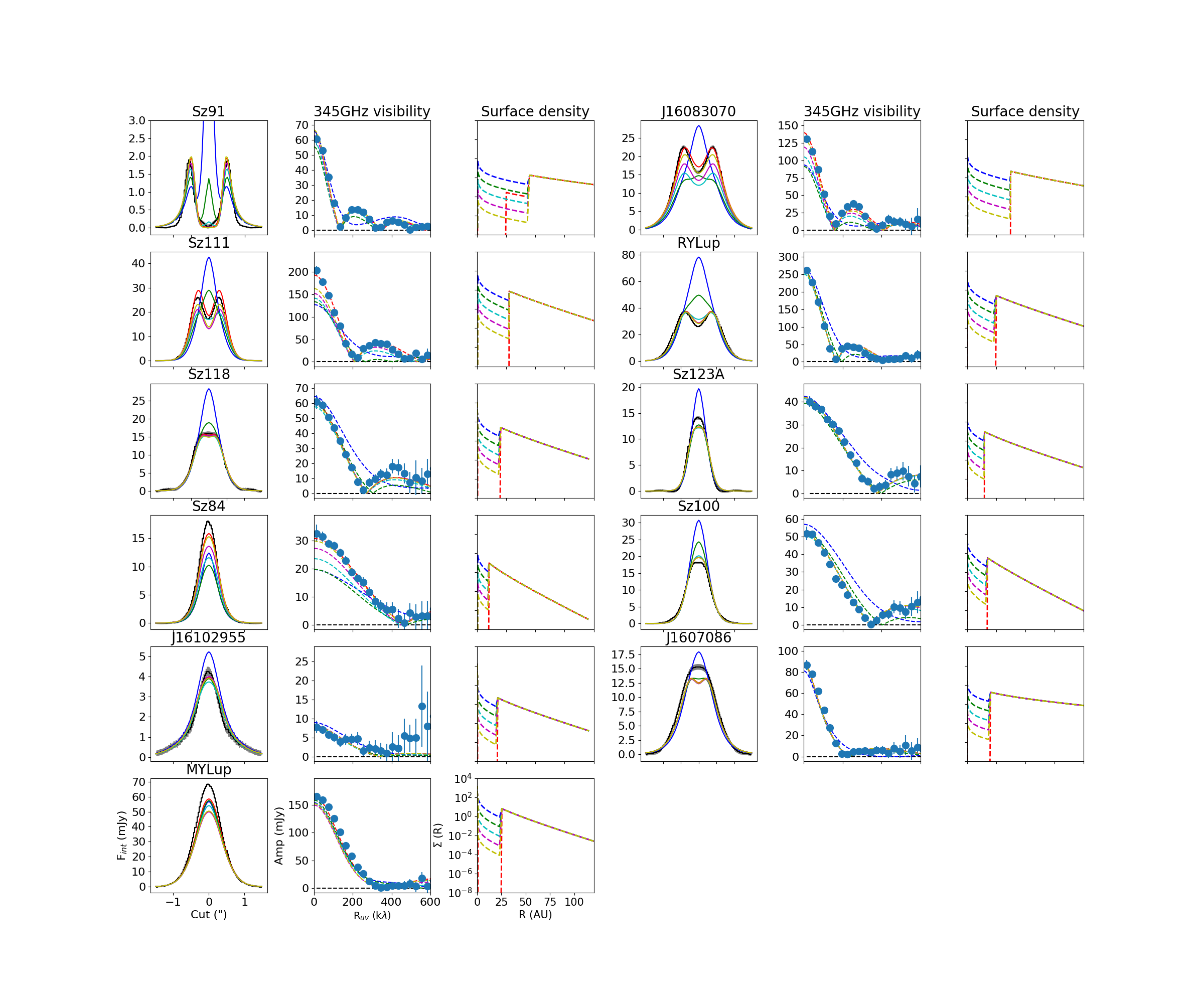}
\caption{Modeling results for the best fit dust model for different amounts of dust inside the dust cavity ($\delta_{\rm dustcav}$). From left to right, for each target, are shown: the azimuthally averaged cut of the 345 GHz continuum; the 345 GHz visibility curve; the density profiles. The data are given in black, the different models in colors green, blue, cyan, purple, yellow and red, for $\delta_{\rm dustcav}$=10$^{-1}$,10$^{-2}$,10$^{-3}$,10$^{-4}$,10$^{-5}$ and 10$^{-20}$. The plots demonstrate that the depth of the dust cavity is at least two orders of magnitude, depending on the target.}
\label{fig:deltadust}
\end{figure*}

\end{document}